%% file: main.tex
\documentclass[
 aip,
 amsmath,amssymb,
reprint
]{revtex4-1}

\usepackage[utf8]{inputenc}
\usepackage[T1]{fontenc}
\usepackage{etoolbox}
\usepackage{graphicx}
\usepackage{dcolumn}
\usepackage{bm}
\usepackage[dvipsnames]{xcolor}
\usepackage{xr-hyper}
\usepackage[colorlinks]{hyperref}
\hypersetup{colorlinks=true,allcolors=Blue}
\usepackage{soul}
\usepackage{xcolor}

\newcommand{\vect}{\boldsymbol}
\newcommand{\mobD}{\Lambda_\mathrm{d}}

\newcommand{\mean}[1]{\langle #1 \rangle}
\newcommand{\figref}[1]{Fig.~\ref{#1}}
\newcommand{\siref}[1]{SI, section~\ref{#1}}
\newcommand{\sifigref}[1]{Supplementary Fig.~\ref{#1}}
\newcommand{\sieqref}[1]{Eq.~\eqref{#1}}
\newcommand{\Eqref}[1]{Eq.~\eqref{#1}}
\newcommand{\Eqsref}[1]{Eqs.~\eqref{#1}}
\newcommand{\cIn}{c_\mathrm{in}}
\newcommand{\cOut}{c_\mathrm{out}}

\newcommand{\vpp}{{\varphi}}

\newcommand{\Freact}{F_\mathrm{react}}

\newcommand{\Vtot}{V_\mathrm{tot}}

\newcommand{\comment}[1]{\iffalse #1 \fi}

\begin{document}

\preprint{APS/123-QED}

\title{Heterogeneous Nucleation and Growth of Sessile Chemically Active Droplets}

\author{Noah Ziethen}
\author{David Zwicker}
 \email{david.zwicker@ds.mpg.de}
\affiliation{
 Max Planck Institute for Dynamics and Self-Organization\\
 Am Faßberg 17 37077 Göttingen
}

\date{\today}

\begin{abstract}
Droplets are essential for spatially controlling biomolecules in cells. To work properly, cells need to control the emergence and morphology of droplets. On the one hand, driven chemical reactions can affect droplets profoundly. For instance, reactions can control how droplets nucleate and how large they grow. On the other hand, droplets coexist with various organelles and other structures inside cells, which could affect their nucleation and morphology. To understand the interplay of these two aspects, we study a continuous field theory of active phase separation. Our numerical simulations reveal that reactions suppress nucleation while attractive walls enhance it. Intriguingly, these two effects are coupled, leading to shapes that deviate substantially from the spherical caps predicted for passive systems. These distortions result from anisotropic fluxes responding to the boundary conditions dictated by the Young-Dupr\'e equation. Interestingly, an electrostatic analogy of chemical reactions confirms these effects. We thus demonstrate how driven chemical reactions affect the emergence and morphology of droplets, which could be crucial for understanding biological cells and improving technical applications, e.g., in chemical engineering.
\end{abstract}

\maketitle

\tableofcontents

\section{Introduction}

Droplets comprised of biomolecules, also known as biomolecular condensates, are crucial for organize biological cells.
For example, such droplets separate molecules, control chemical reactions, and exert forces~\cite{Brangwynne_Eckmann_Courson_Rybarska_Hoege_Gharakhani_Julicher_Hyman_2009,Hyman2014,Banani_Lee_Hyman_Rosen_2017,Dignon2020,Su2021}. 
To fulfill these functions, it is likely that cells control the nucleation, location, size, and shape of droplets.
While nucleation can happen spontaneously inside the cytoplasm~\cite{Shimobayashi2021}, most droplets might be nucleated heterogeneously involving other structures as nucleation sites.
Indeed, many droplets interact with other structures inside cells, like cytoskeletal elements~\cite{Shimobayashi2021,Boddeker2022,Boddeker2023a}, membrane-bound organelles~\cite{Brangwynne_Eckmann_Courson_Rybarska_Hoege_Gharakhani_Julicher_Hyman_2009, Zhao_Zhang_2020, Kusumaatmaja_May_Feeney_McKenna_Mizushima_Frigerio_Knorr_2021}, and  the plasma membrane~\cite{Beutel_Maraspini_Pombo-Garcia_Martin-Lemaitre_Honigmann_2019, Mangiarotti_Chen_Zhao_Lipowsky_Dimova_2023, Pombo-Garcia_Martin-Lemaitre_Honigmann_2022}.
Such interactions of liquid-like droplets with more solid-like structures is known as wetting~\cite{Gennes2004}, and directly linked to heterogeneous nucleation~\cite{Young1805, Cahn_1977, De_Gennes_1981}.
However, most traditional examples of heterogeneous nucleation, e.g., by dust particles in clouds~\cite{Zhang_Khalizov_Wang_Hu_Xu_2012}, concern passive systems.
In contrast, biological cells use external energy input to control processes actively, but it is unclear how activity affects heterogeneous nucleation and the properties of the subsequently forming droplets attached to the solid surface (sessile droplets).

Driven chemical reactions that affect the droplet material are one crucial example for an active process that controls droplets in cells~\cite{Hondele2019, Soeding_Zwicker_Sohrabi-Jahromi_Boehning_Kirschbaum_2019}. 
If such reactions take place in the entire system, droplet size can be controlled~\cite{Soeding_Zwicker_Sohrabi-Jahromi_Boehning_Kirschbaum_2019, Kirschbaum_Zwicker_2021, Zwicker_Hyman_Juelicher_2015}, droplets can divide spontaneously~\cite{Zwicker2017}, and homogeneous nucleation is suppressed~\cite{Ziethen_Kirschbaum_Zwicker_2023}.
Moreover, if reactions are restricted to the boundary, Liese and Zhao et al. recently demonstrated modified shapes of sessile droplets~\cite{liese2023chemically}. 
However, the generic case of wetting in the presence of bulk chemical reactions has not been considered so far.

To study how boundaries affect the behavior of chemically active droplets, we consider a continuous description of phase separation with driven reactions in the bulk and a passive interaction of the droplet material with a flat wall.
Our stochastic simulations~\cite{Cook_1970} and analytical results from an equilibrium surrogate model reveal that reactions generally suppress nucleation, whereas attractive walls facilitate them.
However, both processes exhibit a complex interplay, which is connected to substantial shape deformations:
While critical nuclei maintain their stereotypic spherical cap shapes, macroscopic droplets often exhibit elongated shapes along the boundary.

\section{Model}
\label{sec:model}
We consider an isothermal system of fixed volume $V$ filled with an incompressible, binary mixture of droplet and solvent material with equal molecular volume~$\nu$.
The state of the fluid is described by the concentration field $c(\boldsymbol r, t)$ of the droplet material, whereas the solvent concentration is $\nu^{-1}-c(\mathbf{r}, t)$. 
We describe the interactions and entropy in the system using a free energy comprised of bulk and surface terms given by \cite{Cahn_Hilliard}
\begin{align}
    F=\int_V \left[f(c) + \frac{\kappa}{2}|\nabla c|^2\right]\mathrm{d}\vect r - \int_{\partial V} g(c) \mathrm{d}A
    \;,
  \label{eq:free_en_surf}
\end{align}
where $f(c)$ is the local free energy density, $\kappa$ penalizes compositional gradients, and $g(c)$ is the contact potential describing the interaction of the fluid with the immobile boundary~$\partial V$ of the system.
For simplicity, we focus on the linear order of the expansion $g(c) = g_0 + g_1c + \mathcal{O}(c^2)$ since $g_0$ merely shifts the total free energy $F$, but does not affect the behavior \cite{de_Gennes_1985}.
In contrast, we describe the bulk interactions by 
\begin{align}
    f(c) = 
		-\frac{a_2}{2}\left(c - \frac{1}{2\nu}\right)^2 
		+ \frac{a_4}{4}\left(c - \frac{1}{2\nu}\right)^4 
	\label{eq:f_loc}
	\;,
\end{align}
where $a_2, a_4>0$ are phenomenological coefficients. 

\begin{figure}[t]
    \centering
    \includegraphics[width=0.49\textwidth]{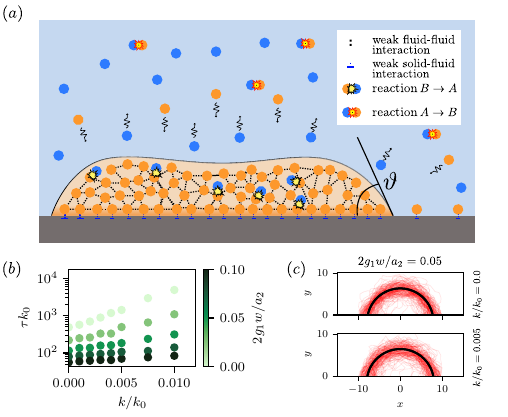}
    \caption{\textbf{Driven reactions delay heterogeneous nucleation.}
    (a) Schematic picture of a sessile reactive droplet. Droplet material is attracted to the wall and itself, leading to phase separation from solvent.
    In addition, active chemical reactions convert droplet material to solvent material inside the droplet, while the converse happens outside.
    This process leads to long-ranged diffusive fluxes, resulting in deformed macroscopic droplets.
    (b) Numerically determined nucleation times $\tau$ as a function of reaction rate $k$ for various wall affinities $g_1$.
    (c) Ensemble of critical nuclei for two reaction rates $k$ and positive wall affinity $g_1=0.025\,a_2/w$. 
    (b-c) Simulations are detailed in the \siref{sec:stoch-sim}; Model parameters are $\nu c_0 =0.1$, $a_2/(\nu k_BT)=200$, $\nu=w^2$, $a_4=4a_2\nu$, and $k_0=\mobD a_2 w^{-2}$.
    }
    \label{fig:shem_and_nucleation}
\end{figure}
Equilibrium states minimize the free energy $F$.
As a necessary condition, the variation of $F$ with respect to $c$ must vanish, which yields the two conditions
\begin{subequations}
\begin{align}
    f'(c) - \kappa \nabla^2 c &= \mathrm{const} && \text{(in the bulk)}
  \label{eq:euler_lagrange}
    \\
    \vect n \cdot \nabla c&=\frac{g_1}{\kappa}&&\text{(at the boundary)} \;,
  \label{eq:boundary_cond}
\end{align}
\end{subequations}
where $\vect n$ denotes the outward normal of the surface $\partial V$.
The first equation describes the balance of the exchange chemical potential $\mu = f'(c) - \kappa \nabla^2 c$, where the constant is determined from material conservation~\cite{Weber_Zwicker_Juelicher_Lee_2019}.
This generically yields a dense phase of concentration $c^{(0)}_\mathrm{in} \approx (2\nu)^{-1} + \sqrt{a_2/a_4}$ that is separated from a dilute phase of concentration $c^{(0)}_\mathrm{out} \approx (2\nu)^{-1} - \sqrt{a_2/a_4}$ by an interface of width~$w=\sqrt{2\kappa/a_2}$. 
The dense phase typically assumes a spherical shape to minimize the surface tension $\gamma_\mathrm{ds}=2 \sqrt{2 \kappa a_2^3}/(3 a_4)$ between the two phases~\cite{Weber_Zwicker_Juelicher_Lee_2019}.
In contrast, \Eqref{eq:boundary_cond} describes a boundary condition, which determines the behavior of the system close to the wall.
In particular, the droplet material is repelled from the wall if $g_1<0$~\cite{Gennes2004}.
Since the droplet still maintains its spherical shape, its geometry at the wall is fully quantified by the contact angle $\vartheta$, which is given by the Young-Dupr\'{e} equation, $\cos(\vartheta)=(\gamma_\mathrm{ws} - \gamma_\mathrm{wd})/\gamma_\mathrm{ds}$ \cite{Young1805}, where $\gamma_\mathrm{ws}$, $\gamma_\mathrm{wd}$, and $\gamma_\mathrm{ds}$ denote the surface tensions between wall-solvent, wall-droplet, and droplet-solvent, respectively.
Since $g(c)$ directly quantifies  surface energies, we have $\gamma_\mathrm{wd}\approx -g_1 c^{(0)}_\mathrm{in}$ and $\gamma_\mathrm{ws} \approx -g_1 c^{(0)}_\mathrm{out}$, resulting in
\begin{align}
    \cos(\vartheta) \approx \frac{3g_1}{a_2}\sqrt{\frac{a_4}{2\kappa}}
    \label{eqn:contact_angle}
    \;.
\end{align}
This equation can only be solved for $\vartheta$ if the interactions between the droplet and the wall are weak, $|g_1|<g_*$ with $g_*\approx\sqrt{(2a_2^2\kappa)/(9a_4)}$.
In contrast, the droplet will be repelled from the wall if $g_1 < -g_*$, and it will fully wet the wall if $g_1 > g_*$.
Since the interesting process of heterogeneous nucleation is related to partial wetting where $\vartheta$ is defined, we concentrate on the case $|g_1|<g_*$.
In particular, we consider the case $g_1>0$, corresponding to an attractive wall that is the most probable site for nucleation.
 
To describe nucleation, we next specify the dynamics of the system.
We start with the continuity equation
\begin{align}
    \partial_t c + \nabla \cdot \vect{j} = s \; ,
    \label{eq:cahn-hilliard-source}
\end{align}
where $\vect j$ denotes the diffusive exchange flux between droplet material and the solvent, and the source term~$s$ describes chemical transitions~\cite{Weber_Zwicker_Juelicher_Lee_2019}.
The passive diffusive flux~$\vect j$ is driven by gradients of the exchange chemical potential, $\vect j = - \mobD \nabla \mu + \vect\eta$, where $\mobD$ is the diffusive mobility and $\vect\eta$ denotes diffusive thermal noise, which obeys $\mean{\eta_i(\vect r, t)}=0$ and the fluctuation dissipation theorem $\mean{\eta_i(\vect r, t) \eta_j (\vect r', t')} = 2 k_{\mathrm{B}} T \mobD \delta_{ij} \delta\left(\vect r - \vect r'\right) \delta\left(t-t'\right)$, where $k_\mathrm{B}T$ is the thermal energy~\cite{Juelicher_Grill_Salbreux_2018,nonEqThermo, Godreche_1991}.
The system becomes active when we drive the reactions described by $s$ out of equilibrium~\cite{Kirschbaum_Zwicker_2021}. 
We focus on driven reactions that result in size-controlled droplets, which requires production of droplet material outside the droplet, while it is degrade inside~\cite{Kirschbaum_Zwicker_2021, Weber_Zwicker_Juelicher_Lee_2019}. 
This behavior can be captured by the linear expression
\begin{align}
    s(c)=-k(c-c_0) \;,
  \label{eq:reaction_flux_lin}
\end{align}
where $k$ sets the reaction rate and $c_0$ denotes the stationary state of the reaction scheme.
We have shown previously that this case faithfully describes homogeneous nucleation in active systems and that the thermal noise associated with the reactions can be neglected since the diffusive noise~$\vect\eta$ dominates on the length scales relevant for nucleation~\cite{Ziethen_Kirschbaum_Zwicker_2023}.
Taken together, our system is described by the stochastic partial differential equation
\begin{align}
	\partial_t c = \mobD \nabla^2 \mu - k(c-c_0) + \nabla \cdot \vect\eta
	\;,
	\label{eqn:pde}
\end{align}
augmented by the no-flux boundary condition $\vect n\cdot \vect j=0$, and \Eqref{eq:boundary_cond} describing local equilibrium at the boundary.

\section{Chemical reactions suppress heterogeneous nucleation}
\subsection{Numerical simulations reveal increased nucleation times}
To investigate heterogeneous nucleation, we performed numerical simulations of \Eqref{eqn:pde} in a two-dimensional system.
Here, we used finite differences to approximate derivatives and an Euler-Maruyama scheme to perform the time stepping~\cite{Zwicker_2020}.
We applied periodic boundary conditions along the $x$-direction and \Eqref{eq:boundary_cond} on both boundaries in the $y$-directions.
Repeating the simulations many times, we observed that the time $t_\mathrm{nucl}$, quantifying when the first droplet nucleates, follows an exponential distribution; see \sifigref{fig:si-histogram}.
We thus define the nucleation time $\tau$ as the ensemble average of $t_\mathrm{nucl}$.
Fig.~\ref{fig:shem_and_nucleation}b shows that $\tau$ decreases for stronger droplet-wall attraction (larger $g_1$), as expected for nucleation of passive droplets \cite{Kalikmanov2013}. 
In contrast, larger reaction rates~$k$ lead to longer nucleation times $\tau$, indicating that active chemical reactions hinder nucleation, consistent with the results for homogeneous nucleation \cite{Ziethen_Kirschbaum_Zwicker_2023}.
These numerical simulations indicate that repulsive walls (small $g_1$) and larger reactions (larger $k$) suppress heterogeneous nucleation, but they do not reveal the underlying principles governing the nucleation process.

\subsection{Equilibrium surrogate model reveals trade-off between wall affinity and chemical reactions}
\label{sec:surr-model}

\begin{figure}[tpb]
    \centering
    \includegraphics[width=0.49\textwidth]{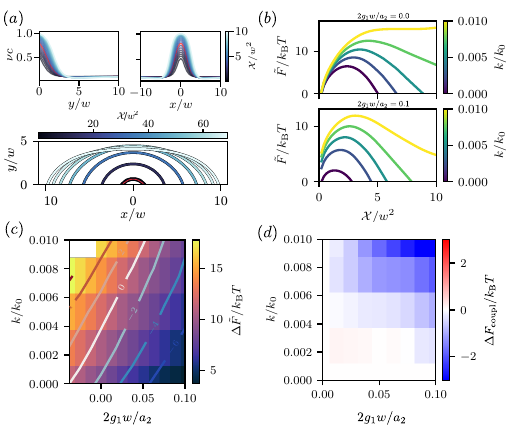}
    \caption{\textbf{Reactions raise energy barrier of nucleation.}
    (a) Cuts through concentration profile $c(x,y)$ along the two axes (top) and shape of droplet interface (bottom) along the nucleation path (colors) for $k/k_0=0.01$ and $g_1w/a_2=0.05$.
    The shape of the critical nucleus is indicated in red. 
    (b) Energy $\tilde{F}$ of the surrogate model as a function of the nucleation coordinate $\mathcal{X}$ for various reaction rates $k$ for a neutral wall ($g_1=0$) and an attractive wall ($g_1=0.05 \,a_2/w$).
    (c) Energy barrier $\Delta \tilde F$ as a function of $g_1$ and $k$.
    (d) Energy $F_\mathrm{coupl}$ following from Eq.~\ref{eq:coupling-ansatz} as a function of $g_1$ and $k$.
    (a-d) Model parameters are as in Fig. \ref{fig:shem_and_nucleation}.}
    \label{fig:path_and_energy}
\end{figure}
To understand the nucleation path in detail, we next map the active system onto a surrogate equilibrium system with long-ranged interactions, which is possible for the special case of the linear reactions that we consider.
Briefly, we can re-write the dynamics given by \Eqref{eqn:pde} as $\partial_t c = \mobD \nabla^2 \delta \tilde F[c]/\delta c + \nabla \cdot \vect \eta$ when we use the augmented free energy functional
\begin{align}
       \tilde F[c] = F[c]  + F_\mathrm{react}[c]
       \;,
       \label{eq:tot_en_func}
\end{align}
where
\begin{align}
	\Freact[c] &= \frac{k}{2 \mobD} \int \bigl[c(\vect r) - c_0\bigr]\Psi(\vect r) \, \mathrm{d}\vect{r}
       \label{eq:energy_react}
\end{align}
captures the energy associated with reactions~\cite{Liu_Goldenfeld_1989, Christensen_Elder_Fogedby_1996, Muratov_2002}. 
Here, $\Psi$ is the solution to the Poisson equation $\nabla^2 \Psi = c_0-c(\vect r)$ and describes the long-ranged interactions, which originate from the interplay of chemical reactions and diffusion in the original model.
Since the surrogate model requires mass conservation, we solve for $\Psi$ employing Neumann boundary conditions, $\vect{n}.\nabla \Psi=0$.

We used the free energy of the surrogate model, given by \Eqref{eq:tot_en_func}, to map out the minimal energy path connecting the homogeneous state with the state where a droplet wets the wall.
To do this, we used constrained optimization to obtain concentration profiles at successively larger values of a nucleation coordinate~$\mathcal{X}$, which measures the amount of material inside the droplet; see \siref{sec:constr-opt}.
\figref{fig:path_and_energy}a shows an example of such a minimal energy path, indicating that the maximal concentration inside the droplet increases alongside its size.

The minimal energy path allows us to evaluate the critical energy and the critical nucleus, which is the transition state of the nucleation process. 
\figref{fig:path_and_energy}b shows that chemical reactions generally increase~$\tilde F$, potentially explaining why reactions suppress nucleation.
However, reactions also increase~$\tilde F$ of the homogeneous state when the wall is attractive ($g_1>0$).
This is because the boundary condition given by \Eqref{eq:boundary_cond} perturbs the homogeneous state in a layer with a thickness of roughly the interfacial width~$w$, leading to local reactive fluxes.
To see how chemical reactions affect nucleation, we thus evaluated the energy difference~$\Delta \tilde F$ between the transition state and the homogeneous initial state.
\figref{fig:path_and_energy}c shows that $\Delta \tilde F$ increases with the reaction rate $k$, while it decreases with more attractive walls (larger $g_1$), which is expected~\cite{Ziethen_Kirschbaum_Zwicker_2023, Kalikmanov2013}.

We next quantified how chemical reactions interact with the wall affinity by decomposing the free energy barrier,
\begin{align}
	\Delta \tilde{F}(g_1, k) = \Delta F_\mathrm{pas}(g_1) + \Delta F_\mathrm{react}(k) + \Delta F_\mathrm{coupl}(g_1, k)
	\;,
  \label{eq:coupling-ansatz}
\end{align}
where $\Delta F_\mathrm{pas} = \Delta \tilde{F}(g_1, k=0)$ quantifies the barrier for passive heterogeneous nucleation, $\Delta F_\mathrm{react} =  \Delta \tilde{F}(g_1=0, k)$ is the energy barrier of chemically driven droplets at a neutral wall, and $\Delta F_\mathrm{coupl}(g_1, k)$ denotes the energy due to the interaction of the two effects.
Since we directly determined $\Delta \tilde{F}(g_1, k)$, $\Delta F_\mathrm{pas}(g_1)$, and $\Delta F_\mathrm{react}(k)$, we can infer $\Delta F_\mathrm{coupl}(g_1, k)$ from \Eqref{eq:coupling-ansatz}.
The data shown in \figref{fig:path_and_energy}d indicates a significant negative coupling between the wall affinity and the reactions, i.e., strong affinity to the wall can decrease the relative effect of chemical reactions.

The coupling between the wall affinity~$g_1$ and the reaction rate~$k$ is also apparent in the contour lines of equal~$\Delta \tilde F$, where reactions and wall affinity compensate each other; see \figref{fig:path_and_energy}c.
We show in the \siref{sec:coupling} that concave contour lines would be expected without coupling ($\Delta F_\mathrm{coupl}=0$), so that the slight convex shape of the observed lines is a strong indication for coupling.
To analyze these contour lines in more detail, we approximate the coupling by a bilinear function, $\Delta F_\mathrm{coupl}(g_1, k)=hg_1k$ with pre-factor $h$, motivated by \figref{fig:path_and_energy}d.
Moreover, we use a linear approximation for the reactive energy~\cite{Ziethen_Kirschbaum_Zwicker_2023}, $\Delta F_\mathrm{react}=mk$, and we express the energy associated with the passive case as \cite{Kalikmanov2013}
\begin{align}
    \Delta F_\mathrm{pas}(g_1) 
    &= \frac{\gamma_\mathrm{ds}^2}{ \Delta f} \bigl[2\vartheta(g_1) - \sin(2\vartheta(g_1))\bigr]
    \;,
  \label{eq:delta-F-pas}
\end{align}
where $\Delta f=f(c_0)-f(c_\mathrm{in}^\mathrm{eq})+f'(c_0)(c_\mathrm{in}^\mathrm{eq}-c_0)$ and $\vartheta$ is given by \Eqref{eqn:contact_angle}. 
Taken together, the contour line associated with energy $\Delta F_\mathrm{c}$ then satisfies
\begin{align}
	k = \frac{1}{m + hg_1} \left[ \Delta F_\mathrm{c} - \Delta F_\mathrm{pas}(g_1)\right]
	\;. 
  	\label{eq:contour-line}
\end{align}
We show in the \siref{sec:coupling} that the curvature of these contour lines is typically negative ($\partial^2 k/\partial g_1^2>0$), consistent with the observed convex shape.

To explore the observed coupling further, we next ask how the energy $F_\mathrm{react}$, defined by \Eqref{eq:energy_react}, changes with the wall affinity~$g_1$.
We thus initialized droplets of fixed volume for various affinities~$g_1$ and numerically calculated $F_\mathrm{react}$ inside the droplet; see \siref{sec:react-energy-inside}. 
\sifigref{fig:si-coupling-3} shows that $F_\mathrm{react}$ indeed decreases with larger $g_1$, consistent with the negative effect we found above.
However, this analysis only probes how different contact angles affect the reactions, whereas the reactions potentially also change the entire droplet shape away from a spherical cap.

\section{Sessile active droplets spread along walls}
\subsection{Droplets deviate from spherical cap shape after nucleation}
The minimal energy paths that we obtained from the surrogate model  not only provide energy barriers for nucleation but also the most likely droplet shape along the nucleation path.
\figref{fig:path_and_energy}a shows that small droplets are shaped like a spherical cap, which corroborates the sampled critical droplet shapes shown in \figref{fig:shem_and_nucleation}c.
However, \figref{fig:path_and_energy}a also shows that larger droplets can deviate significantly from this equilibrium shape.

The droplet shape originates from a minimization of the free energy $\tilde{F}$ of the surrogate equilibrium model. 
If reactions are absent, surface tension~$\gamma_\mathrm{ds}$ minimizes the interface between droplet and solvent material, resulting in a spherical shape with constant mean curvature.
At the systems boundary, the shape must be compatible with the contact angle~$\vartheta$ controlled by the energy balance that led to \Eqref{eqn:contact_angle}, which implies that attractive walls result in flatter spherical caps.
The situation is more complicated with chemical reactions.
The long-ranged interaction described by \Eqref{eq:energy_react} leads to a repulsion of the droplet material, analogously to electrostatic repulsion of charged material.
Consequently, different parts of the droplet repel each other~\cite{Rayleigh1882}, which can induce spontaneous splitting for large bulk droplets~\cite{Golestanian2017, Zwicker2017} and explains why active droplets spread along the wall.
\begin{figure}
    \centering
    \includegraphics[width=0.5\textwidth]{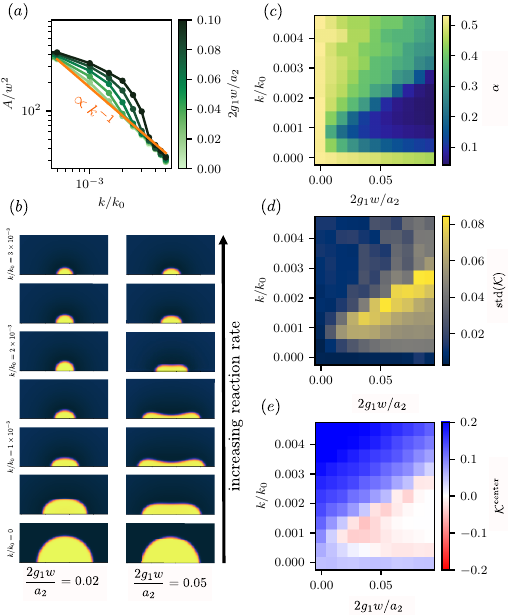}
    \caption{\textbf{Active droplets spread along walls.}
    (a)~Droplet area as a function of the reaction rate $k$ for different wall affinities $g_1$.
    (b) Concentration fields of sessile droplets for two values of $g_1$ and various $k$.
    (c) Aspect ratio $\alpha$ of sessile droplet as a function of $g_1$ and $k$. 
    (d) Standard deviation of the interface curvature $\mathcal{K}$ as a function of $g_1$ and $k$.
    (e) $\mathcal{K}$ evaluated at the droplet center as a function of $g_1$ and $k$.
    (a--e) Model parameters given in Fig.~\ref{fig:shem_and_nucleation}.
    }
    \label{fig:equilibrium_droplets}
\end{figure}

\subsection{Interplay of reactions and wall affinity can deform macroscopic droplets}
To understand how the interplay between wall affinity and driven chemical reaction influences the droplet shape, we next  study the stationary
shapes of sessile droplets numerically.
We initialized droplets at the wall and simulated their time evolution for different reaction rates $k$ and wall affinities $g_1$ until steady states were reached. 
To get an initial understanding, we first quantified droplet size, which is an area in our two-dimensional simulations.
If droplet size is controlled by the reaction-diffusion length scale $L=(D/k)^{1/2}$, with some diffusivity~$D$, we expect the droplet size to scale with $k^{-1}$.
\figref{fig:equilibrium_droplets}a indeed reveals such a scaling, at least for large reaction rates.
In contrast, for intermediate rates, we observe significant deviations, which increase with the wall affinity.
The snapshots shown in \figref{fig:equilibrium_droplets}b suggest that intermediate reaction rates~$k$ lead to strongly deformed droplets, which could explain the observed size-dependence.

We next quantified droplet shapes by analyzing the curve describing the interface, defined as the iso-contour $c=0.5$.
First, we determined the aspect ratio~$\alpha$, which is defined as the quotient of the lengths of the minor and major axis (see \figref{fig:equilibrium_droplets}c).
Second, we quantified the curvature $\mathcal{K}$ of the interface (see \siref{sec:calc-curv}) and determined its variation along the interface (see \figref{fig:equilibrium_droplets}d) as well as the curvature at the center point (see \figref{fig:equilibrium_droplets}e).
All three quantifications shown reveal the same fundamental dependence:
Droplets are essentially spherical caps when reactions are absent ($k=0$), consistent with expectations~\cite{Gennes2004}.
Moreover, droplets are spherical for neutral walls ($g_1=0$), even when reactions are strong, because all fields are radially symmetric in this symmetric case.
Interestingly, droplets also exhibit a spherical cap shape for large reaction rates, likely because strong reactions reduce droplet size so interfacial effects dominate. 
Between these extreme values, we observe strong deviations from a spherical shape, both in the snapshots (\figref{fig:equilibrium_droplets}b) and in the quantifications (\figref{fig:equilibrium_droplets}c--e). 
In fact, for a given wall attraction~$g_1$, we empirically find an intermediate rate~$k$ for which the deformations are maximally, and this rate increases with $g_1$.
Taken together, these quantifications reveal an interplay between wall attraction and reactions, which leads to macroscopic deformations of sessile droplets.

\begin{figure}[t]
    \centering
    \includegraphics[width=0.45\textwidth]{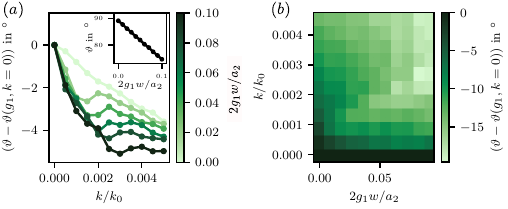}
    \caption{\textbf{Reactions affect contact angles.}
    (a,~b)~Deviation of contact angle $\vartheta$ from passive case ($k=0$) as a function of reaction rate $k$ for various wall affinities $g_1$.
    The inset shows $\vartheta$ as a function of $g_1$ for $k=0$.
	Model parameters given in Fig. \ref{fig:shem_and_nucleation}.}
    \label{fig:wetting-angles}
\end{figure}
The macroscopic deformations induced by the reactions affect the entire interface and could thus also impact the contact angle~$\vartheta$.
To test this, we evaluated the slope $-\partial_x c/\partial_y c$ along the implicit curve $c(x_I,y_I)=0.5$ and measured the associated $\vartheta$ at the boundary.
The inset in \figref{fig:wetting-angles}a shows that $\vartheta$ decreases with increasing wall affinity~$g_1$, consistent with the prediction from \Eqref{eqn:contact_angle}.
However, the data in \figref{fig:wetting-angles}a also indicates that $\vartheta$ decreases for larger reaction rates~$k$, indicating an effect of the reactions.
For attractive walls ($g_1>0$), the contact angle first declines sharply for increasing $k$, although this decline becomes weaker for larger $k$ and there is even a brief non-monotonic behavior.
The value of $k$, where this non-monotonic behavior is observed, depends on $g_1$ and coincides with the maximal shape deformation of the droplets (compare \figref{fig:wetting-angles}b to \figref{fig:equilibrium_droplets}a).
Chemically active sessile droplets thus exhibit changes in apparent contact angle concomitantly with global shape deformations.

\subsection{Anisotropic fluxes cause  droplet deformation}
To gain further insights into shape deformations, we next analyze how reactions disturb the spherical cap shapes expected for passive droplets. 
We consider a spherical cap described by a radius~$R$ and contact angle~$\vartheta$ in the half-space $y>0$. 
To describe the droplet shape explicitly, we use an effective droplet model~\cite{Weber_Zwicker_Juelicher_Lee_2019}, assuming a thin interface ($w \ll R$), so we can approximate the dynamics of the concentration fields~$c_i$ by reaction-diffusion equations
\begin{align}
    \partial_t c_i \approx D \nabla^2 c_i - k (c_i - c_0)
    \;,
  \label{eq:reaction_diffusion}
\end{align}
inside ($i=\mathrm{in}$) and outside ($i=\mathrm{out}$) the droplet; see \siref{sec:eff-drop}.
Here, we linearized \Eqref{eqn:pde} without noise around the concentrations $c^{(0)}_\mathrm{in}$ and $c^{(0)}_\mathrm{out}$, so the diffusivity is given by $D=\mobD f'{'}(c^{(0)}_\mathrm{in})=\mobD f'{'}(c^{(0)}_\mathrm{out})$ for our choice of a symmetric free energy~\cite{Weber_Zwicker_Juelicher_Lee_2019}.
To solve the reaction-diffusion problem, we employ polar coordinates $(r, \varphi)$ centered at the sphere describing the spherical cap. 
We impose $\partial_y c_i|_{y=0} = 0$ and $\partial_r c_\mathrm{out}|_{r\rightarrow \infty}=0$ at the system boundary.
The conditions at the droplet interface are governed by the local phase equilibrium and read $c_i(R)=c_i^{(0)}+\beta_i \gamma_\mathrm{ds}/R$, where $\beta_i=2/[(c_\mathrm{in}^{(0)}-c_\mathrm{out}^{(0)})f''(c_i^{(0)})]$ quantifies surface tension effects for $i=\mathrm{in}, \mathrm{out}$~\cite{Weber_Zwicker_Juelicher_Lee_2019}. 
For simplicity, we consider the quasi-stationary situation where the concentration fields equilibrate faster than the interfacial shape.
The resulting stationary state solutions to \Eqref{eq:reaction_diffusion} then depend on the reaction-diffusion length scale $L= \sqrt{D/k}$, the average concentration $c_0$, and the boundary conditions. 
The general solution of this Helmholz equation reads
\begin{multline}
    c_i(r, {\varphi})=
   c_0\\+\sum_{n=0}^{\infty}\!\left[A_n^i I_{\lambda_n}\!\!\left(\frac{r}{L}\right)+B_n^i K_{\lambda_n}\!\!\left(\frac{r}{L}\right)\right]\! 
   \cos (\lambda_n\bigl({\varphi}-\vartheta)\bigr)
    \;,
  \label{eq:hemholz-sol}
\end{multline}
where $I_{\lambda_n}(z)$ and $K_{\lambda_n}(z)$ are the modified Bessel functions of first and second kind.
Here, the wavelength $\lambda_n=n\pi/\vartheta$ of the polar coordinate is quantized by the boundary conditions, which can also be used to determine the series coefficients $A_n^i$ and $B_n^i$ as described in the \siref{sec:eff-drop}.
Taken together, this approach provides approximate solutions for the concentration profiles inside and outside the droplet.

The concentration fields imply fluxes $\vect j _i = -D \nabla c_i$, which can affect droplet shapes.
Indeed, the interface speed ${v}_\mathrm{n}$ normal to the interface reads~\cite{Weber_Zwicker_Juelicher_Lee_2019}
\begin{align}
  {v}_\mathrm{n}(\varphi) =
  	\frac{\boldsymbol{j}_{\mathrm{in}}(r=R, \varphi)-\boldsymbol{j}_{\mathrm{out}}(r=R, \varphi)}
    {c_{\mathrm{in}}^{(0)}-c_{\mathrm{out}}^{(0)}}\cdot \vect{n}
  \;,
  \label{eq:interface-vel}
\end{align}
where $\vect{n}$ is the normal vector of the interface.
Using \Eqref{eq:hemholz-sol}, we find
\begin{align}
    {v}_\mathrm{n}  =\frac{1}{c_{\mathrm{in}}^{(0)}-c_{\mathrm{out}}^{(0)}} 
    	\sum_{n=0}^\infty \mathcal{C}_n(R, L, \vartheta) \cos \left(\lambda_n {\varphi}\right)
   \;,
  \label{eq:interface_speed}
\end{align}
with
\begin{multline}
\mathcal{C}_n(R, L, \vartheta)=\left(D_{\text {out }} A_n^{\text {out }}-D_{\text {in }} A_n^{\text {in }}\right) I_{\lambda_n}^{\prime}\left(\frac{R}{L}\right)
\\
	+\left(D_{\text {out }} B_n^{\text {out }}-D_{\text {in }} B_n^{\text {in }}\right) K_{\lambda_n}^{\prime}\left(\frac{R}{L}\right).
\end{multline}
This interfacial speed~${v}_\mathrm{n}$ quantifies how chemical reactions would disturb the spherical cap shape.

\begin{figure}[t]
    \centering
    \includegraphics[width=0.48\textwidth]{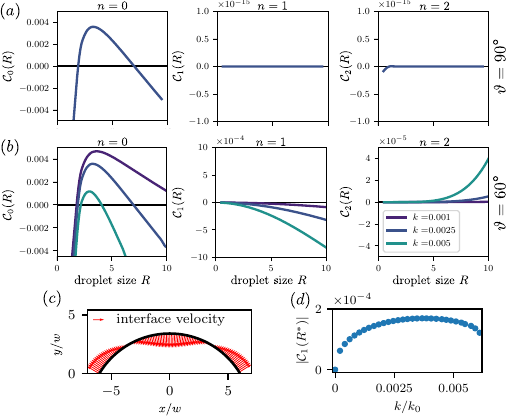}
    \caption{\textbf{Reactions cause non-isotropic interface speeds.}
    (a) Series coefficients $\mathcal{C}_n$ as a function of droplet size~$R$ for $n=0,1,2$ at a neutral wall ($g_1=0$, contact angle $\vartheta=90^\circ$) for $k=0.0025$.
    (b) $\mathcal{C}_n$ as a function of $R$ for an attractive wall ($\vartheta=60^\circ$).
    (c) Interface velocity $v_\mathrm{n} \vect{n}$ (red arrows) predicted from Eq.~\eqref{eq:interface_speed} using the first three modes for a spherical cap of $R/w=7$ for $\vartheta = 60^\circ$ (black line).  
    (d) $|\mathcal{C}_1|$ as a function of the reaction rate $k$ for $\vartheta=60^\circ$.
    (a--d) Additional parameters  are derived from values given in Fig. \ref{fig:shem_and_nucleation}.}
    \label{fig:interface_coefficents}
\end{figure}
We start by examining the shape of a droplet on a neutral wall ($g_1=0$). 
\figref{fig:interface_coefficents}a shows that only the $n=0$ mode contributes, whereas $A^i_n=B^i_n=0$ for $n\ge1$, consistent with the observed spherical shapes.
The zeroth mode is essentially unchanged for an attractive wall ($g_1>0$, $\vartheta<\pi/2$, \figref{fig:interface_coefficents}b), but the first and second modes become important for larger droplet sizes, indicating non-spherical droplet shapes.
The dependence of these modes on the droplet radius~$R$ captures the behavior we observed so far:
For very small radii, the zeroth mode is negative, indicating an unstable size.
Consequently, droplets can only grow spontaneously when $R$ exceed the first root of $\mathcal{C}_0(R)$, which thus corresponds to the size of the critical nucleus.
At the critical size, the higher modes shown in the right two panels in \figref{fig:interface_coefficents}b are vanishingly small, consistent with the spherical shape of critical nuclei observed in \figref{fig:shem_and_nucleation}c.
Droplets larger than the critical size grow until they reach a stationary state, marked by the second root of $\mathcal{C}_0(R)$ in \figref{fig:interface_coefficents}b; see also ref.~\cite{Zwicker_Hyman_Juelicher_2015}.
At this stationary state, the zeroth mode does not contribute to the dynamics, but higher-order modes, which are connected to shape deformations, do.
Indeed, \figref{fig:interface_coefficents}c shows that the interface speed of a stationary droplet with radius $R^*$, chosen such that $\mathcal{C}_0(R^*)=0$, is such that the droplet flattens and spreads along the walls. 
Taken together, the analysis of the first mode of a droplet with a stationary size indicates that fluxes caused by the driven reactions deform the droplet.

To understand droplet deformations in detail, we next focus on the first mode ($n=1$).
\figref{fig:interface_coefficents}d shows that its magnitude $|\mathcal{C}_1(R=R^*, k)|$ vanishes when reaction are absent ($k=0$), consistent with the expected spherical cap shape of passive droplets.
Larger reaction rates~$k$ first cause $|\mathcal{C}_1|$ to increase, but $|\mathcal{C}_1|$ then decreases beyond a critical value of $k$.
This non-monotonic behavior is qualitatively consistent with the shape deformations shown in Fig. \ref{fig:equilibrium_droplets} and might be related to the steady state size of reactive droplets:
Higher reaction rates~$k$ reduce the droplet size (see \figref{fig:equilibrium_droplets}a and ref.~\cite{Zwicker_Hyman_Juelicher_2015}), leading to smaller magnitudes of the droplet deforming modes; see \figref{fig:interface_coefficents}b. 
In summary, chemical reactions combined with symmetry breaking by an attractive wall lead to non-isotropic flows that cause droplet deformation.

\subsection{Reactions deform sessile 3D droplets}
\begin{figure}
    \centering
    \includegraphics[width=0.48\textwidth]{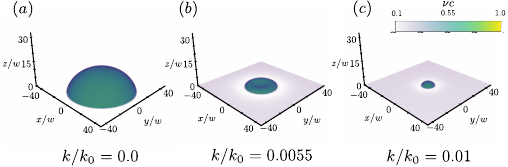}
    \caption{\textbf{Reactions deform sessile 3D droplets.}
    Concentration profiles $c(x, y, z)$ of 3D droplets in stationary state for various reaction rates $k$ at an attractive wall ($g_1=0.05\, a_2/w$).
Simulations were performed in cylindrical symmetry using the parameters $r_\mathrm{max}/w=64$, $dr/w=1$, $z_\mathrm{max}/w=32$, $dz/w=0.25$. Additional parameters can be found in \figref{fig:shem_and_nucleation}.}
    \label{fig:3d}
\end{figure}

So far, we analyzed deformed droplets only in two spatial dimensions.
To check whether the observed effects persist in three dimensions, we performed a simulation for the parameter regime where droplets are deformed.
\sifigref{fig:si-3d} shows that such droplets are circular in the $xy$-plane and are deformed only in the $z$-direction.
Consequently, simulations using cylindrical symmetry are suitable to describe the problem.
\figref{fig:3d} shows that we find spherical cap shapes for small and large reaction rates~$k$, and strongly deformed droplets between these extremes, consistent with our results for attractive walls in two-dimensional systems.
Taken together, we thus expect that our results translate directly to three-dimensional systems.

\section{Electrostatic analogy explains effects of driven reactions}
So far, we have focused on simulating the dynamical equation \eqref{eqn:pde} and analyzed the simplified reaction-diffusion equation \eqref{eq:reaction_diffusion}.
In this final section, we now investigate the equilibrium surrogate model given by \Eqref{eq:tot_en_func} in detail, which will allow us to interpret the reactions as long-ranged electrostatic interactions.

\begin{figure}
    \centering
    \includegraphics[width=0.48\textwidth]{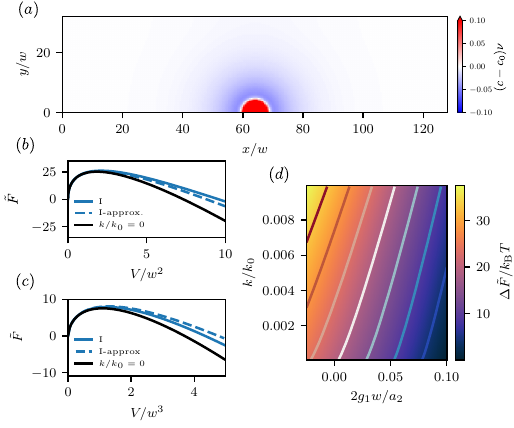}
    \caption{\textbf{Electrostatic analogy explains nucleation behavior.}
    (a) Effective charge density $c(\vect r) - c_0$ of a sessile droplet for reaction rate $k/k_0=0.005$. 
    (b, c) Energy barrier~$\tilde F$ as a function of the volume $V$ in 2D (panel b, $k/k_0=0.001$) and 3D (panel c, $k/k_0=0.01$) for various approximations:
    The full theory (solid blue lines, see \sieqref{eq:energy-total-2d} and \sieqref{eq:energy-total-3d}) is compared to the approximations given in \Eqsref{eq:energy-single-drop} (dashed lines) and the expression for $k=0$ (black lines). 
    The contact angle is $\vartheta=80^\circ$. 
(d) Nucleation barrier~$\Delta \tilde F$ determined from maximizing $\tilde{F}(V)$ given by \Eqref{eq:energy-single-drop-2d} for various $k$ and $g_1$.
(a--d) Additional parameters are given in \figref{fig:shem_and_nucleation}.}
    \label{fig:elect-stat-1}
\end{figure}

The energy~$\Freact$ associated with reactions, given by \Eqref{eq:energy_react}, reveals that the deviation of the concentration field $c(\vect r)$ from the reaction equilibrium $c_0$ can be interpreted as a charge density.
This interpretation only works for the linearized reactions given by \Eqref{eq:reaction_flux_lin} and when the average concentration field is equal to $c_0$, e.g., when $\int c \, \mathrm{d} V = c_0 V$, which implies a charge neutral system in the electrostatic interpretation.
\figref{fig:elect-stat-1}(a) shows a typical concentration profile of a sessile droplet, indicating that the droplet can be interpreted as a positively charged ball  surrounded by a cloud of negative charges, so that the entire system is charge neutral.
The reaction-diffusion length scale~$L= \sqrt{D/k}$ controls the extent of the cloud and the reaction rate~$k$ determines the magnitude of the electrostatic interaction.
Even though the sessile droplet in \figref{fig:elect-stat-1}(a) looks as if it would form a dipole with the surrounding negative charges, this is not the case since image charges restore the symmetry.
Distant droplets thus hardly interact with each other.

To be more quantitative, we next calculate the total energy of a single sessile droplet in the limit of a thin interface, also known as a capillary approximation; see \siref{sec:scaling}.
The approximate expressions for small droplets in two and three dimensions read \cite{Ziethen_Kirschbaum_Zwicker_2023,Muratov_2002}
\begin{subequations}
\label{eq:energy-single-drop}
\begin{align}
    \tilde{F}_\mathrm{2D} &\approx
	\bigl[0.029+0.32\log(LV_\mathrm{2D}^{-\frac12})\bigr]\frac{(\cIn^{(0)}-c_0)^2}{\mobD}kV_\mathrm{2D}^2
\notag\\ &\quad
    -\Delta f V_\mathrm{2D} + \gamma_\mathrm{ds} A_\mathrm{2D}
    \;,
    \label{eq:energy-single-drop-2d}
\\
    \tilde{F}_\mathrm{3D} &\approx
    0.039\frac{(\cIn^{(0)}-c_0)^2}{\mobD} kV_\mathrm{3D}^{\frac53}
    -\Delta f V_\mathrm{3D} + \gamma_\mathrm{ds} A_\mathrm{3D} 
    \;, 
  \label{eq:energy-single-drop-3d}
\end{align}
\end{subequations}
where the bulk energy proportional to $\Delta f=f(c_0)-f(\cIn) + f'(c_0)(\cIn-c_0)$ scales with the droplet volume~$V$, whereas the surface energy proportional to $\gamma_\mathrm{ds}$ scales with the effective size of the interface, which depends on the contact angle $\vartheta$:
$A_\mathrm{2D}=\sqrt{2V_\mathrm{2D}(2\vartheta -\sin(2\vartheta))}$ and 
$A_\mathrm{3D}=4\pi(3V_\mathrm{3D}/[\pi(2 + \cos(\vartheta))(1-\cos(\frac\vartheta2)^2)])^{\frac23}\sin(\frac\vartheta2)^4$.
For simplicity, we neglect the influence of $\vartheta$ on the electrostatic energies associated with the charge distribution given by the respective first terms proportional to $k$.
Instead, we use the expression for one half of a spherical droplet on a wall as an approximation; see \siref{sec:scaling}. 
Neglecting logarithmic corrections, we thus find that reactive energies scale as $R^{2+d}$, bulk energies as $R^d$, and surface energies as $R^{d-1}$, when droplets are small and $d$ is the space dimension.

We first use the approximate energies to investigate nucleation.
The surface energy dominates for small droplets, explaining why nucleated droplets close to the critical size are spherical; see \figref{fig:shem_and_nucleation}c.
\Eqsref{eq:energy-single-drop} also show that chemical reactions raise the nucleation barrier (see \figref{fig:elect-stat-1}(b,c)), consistent with suppressed nucleation.
The associated energy barrier~$\Delta\tilde F$ shown in \figref{fig:elect-stat-1}d exhibits a very similar dependence on the reaction rate $k$ and the wall affinity $g_1$ 
compared to the numerically determined barriers shown in \figref{fig:path_and_energy}c. 
However, since we do not consider the contact angle~$\vartheta$ in the simplified electrostatic energy given in \Eqref{eq:energy-single-drop}, this approximation cannot capture the coupling between $k$ and $g_1$ we revealed above.
Moreover, the capillary approximation used here is known to overestimates the energy barrier for droplet formation~\cite{Kalikmanov2013} since it assumes a well-defined separation of the droplet and the surrounding dilute phase during nucleation.
Yet, we can conclude that the electrostatics do not affect the shape of small droplets, but they oppose their formation, thus lowering nucleation rates.

\begin{figure}
    \centering
    \includegraphics[width=0.48\textwidth]{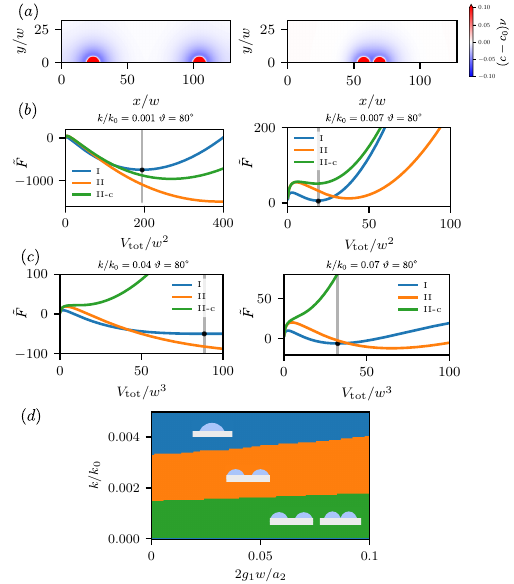}
    \caption{\textbf{Electrostatic analogy explains droplet deformations.}
    (a) Effective charge density $c(\vect r) - c_0$ of two sessile droplet that are well-separated (left) or directly adjacent (right) for for a reaction rate $k/k_0=0.005$. 
    (b, c) Energy barrier~$\tilde F$ as a function of the total volume $\Vtot$ of one droplet (I, blue lines) and two droplets that are well-separated (II, orange lines) or adjacent (II-c, green lines) in 2D (b) and 3D (c) determined using the full theory; see ~\siref{sec:scaling}.
    Parameters are indicated in the panels. 
    (d) State diagram indicating which state has the lowest energy~$\tilde F$ at the total volume where a single droplet is stable as a function of reaction rate $k$ and wall affinity $g_1$.
     The single droplet has the lowest energy in the blue region and for $k=0$, whereas the separated droplets have lower energy in the orange region, and both connected and separated droplets have lower energies in the green region.
 (a--d) Additional parameters are given in \figref{fig:shem_and_nucleation}.}
    \label{fig:elect-stat}
\end{figure}
To discuss shape deformations and splitting of droplets, we next approximate these two situations by considering two sessile droplets with total volume $\Vtot$ that either touch each other or are well-separated; see \figref{fig:elect-stat}a.
If $\tilde F(\Vtot)$ denotes the energy of the single droplet of volume $\Vtot$, the two well-separated droplets have a total energy of $2\tilde F(\frac12 \Vtot)$.
In addition, the two connected droplets exhibit an electrostatic repulsion, which we approximate as the repulsion between two point charges; see \siref{sec:scaling}. 
Since the approximations given in \Eqsref{eq:energy-single-drop} only hold for small droplets, we here obtain $\tilde F(V)$ from \sieqref{eq:energy-total-2d} and \sieqref{eq:energy-total-3d} in the Supplement.
Taken together, we can thus compare the energies of the two droplets at different separation to the energy of a single droplet in 2D and 3D for various $\Vtot$; see \figref{fig:elect-stat}(b,c).
Without reactions ($k=0$), the energy of two droplets is always higher than that of a single droplet, consistent with a minimization of surface energies and absent droplet deformations.
Moreover, $\tilde F(V)$ decreases monotonously beyond the critical size, implying that droplets grow until they are limited by system size.
The picture changes qualitatively when reactions are enabled (and droplets become charged in the electrostatic picture):
Even for arbitrarily small $k$, $\tilde F(V)$ develops a minimum at a finite size $V$ and the electrostatic energy makes larger droplets unfavorable; see \figref{fig:elect-stat}(b,c).
Since the minimal energy is negative (implying that the droplet is favored over the homogeneous state), we observe that the states with two droplets (orange curves) have lower minima than the one-droplet states (blue curves), which implies that a large system would prefer to have many droplets.
A large droplet count could emerge either from nucleating additional droplets or from splitting existing droplets.

To investigate droplet splitting, we compare the minimal energy of a single droplet with the energy of the two-droplet states at the corresponding total volume~$\Vtot$.
\figref{fig:elect-stat}b shows that the two-droplet states can have lower energy for small reaction rates $k$, suggesting that a droplet with this volume might deform spontaneously. 
In contrast, the right panel shows that a single droplet has lower energy for large reaction rates $k$, suggesting that the single droplet is stable.
The transition between these cases marks the onset of droplet deformation, which depends on the reaction rate~$k$ and the wall affinity~$g_1$.
\figref{fig:elect-stat}d suggests that droplets deform if reactions are not too strong and larger $g_1$ favor deformations; similar to \figref{fig:equilibrium_droplets}(c-e).
However, in contrast to \figref{fig:equilibrium_droplets}(c-e), the simple scaling theory predicts that the deformed droplets (mimicked by the two adjacent droplets) are favored even when $g_1=0$.
It is likely that deformations are kinetically suppressed, i.e., that small deformations are energetically unfavorable and we thus do not observe them.

The predicted transition from a single to a deformed and then split droplet  can be understood as a competition between surface energies (favoring compact droplets) and effective electrostatic repulsion (favoring elongated droplets) in conjunction with the favored finite droplet size due to reactions. 
Droplets are spherical at large reaction rates~$k$ since large $k$ implies small droplets where surface energies dominate.
Smaller reactions imply larger droplets, where the electrostatic effects are stronger and eventually favor deformed droplets that can also split.
However, if reactions are absent, we again observe spherical droplets since there are no electrostatic effects.

In summary, interpreting the reactions using electrostatics explains the delayed nucleation and droplet deformation.
Nucleation is suppressed since electrostatics disfavor an accumulation of charges, but the critical droplets are still spherical since surface energies dominate.
In contrast, larger droplet can minimize the total energy by deforming, which enlarges the average distance between charges and thus lowers the electrostatic energy at the expense of a larger surface energy.
In this case, the deformed droplets can also grow beyond the volume predicted for spherical shapes, consistent with \figref{fig:equilibrium_droplets}a.

\section{Discussion}
We showed that nucleation is accelerated by attractive walls and suppressed by active chemical reactions, consistent with our previous results~\cite{Ziethen_Kirschbaum_Zwicker_2023}.
However, we also found an intricate interplay between the two processes, likely stemming from shape modifications in sessile droplets.
While the shape of critical nuclei is in agreement with a spherical cap, consistent with passive scenarios~\cite{Gennes2004}, we found massive deformations for larger droplets at intermediate reaction rates. 
Within the surrogate model, the elongated droplets can also be interpreted as a trade-off between surface tension and effective electrostatic repulsion, which demonstrates that the active chemical reactions mediate a repulsive long-ranged interaction.

Our study of sessile active droplets provides the first step toward understanding the interplay of chemically active droplets with other structures.
For simplicity, we focused on a two-component mixture, whereas cells exhibit a staggering complexity involving thousands of different components, which could be described by a multicomponent extension of our theory~\cite{Mao2018, Zwicker_Laan_2022}. 
Furthermore, we considered reactions that depend linearly on composition, but realistic, thermodynamically consistent reactions are more complex~\cite{Zwicker2022a}.
While our previous work suggests that linear reactions capture nucleation quantitatively~\cite{Ziethen_Kirschbaum_Zwicker_2023}, the shape of macroscopic droplets might change drastically, e.g., when surface reactions are additionally taken into account~\cite{liese2023chemically}.
Finally, we only studied flat walls, but boundaries in cells are typically curved and deformable~\cite{Agudo-Canalejo_Schultz_Chino_Migliano_Saito_Koyama-Honda_Stenmark_Brech_May_Mizushima_2021, Kusumaatmaja_May_Feeney_McKenna_Mizushima_Frigerio_Knorr_2021}.
Incorporating these aspects will likely require more advanced computational methods (such as~\cite{Mokbel2023}), but we expect the general behavior that we unveiled here to persist.

Active control of phase separation is a challenge in cells. 
Our work suggests that cells could use chemical reactions to fine-tune the rate of heterogeneous nucleation as well as the shape of sessile droplets. 
Such deformed droplets may offer better control of wall deformations by condensation, e.g., by using less material than a spherical cap to achieve the same deformation.
Such a regulation process would be a fascinating example combining control via active chemistry and a membrane surface~\cite{Snead2019}.
Moreover, our model serves as an intriguing example of boundary effects in active field theories, inviting comparisons with other active field theories~\cite{Turci_Wilding_2021, Turci_Jack_Wilding_2023}.

\begin{acknowledgments}
We thank Yicheng Qiang and Frieder Johannsen for helpful discussions and Cathelijne ter Burg for critical reading of the manuscript.
We gratefully acknowledge funding from the Max Planck Society, the German Research Foundation (DFG, grant agreement ZW 222/3), and the European Union (ERC, EmulSim, 101044662).
\end{acknowledgments}

\section*{Data Availability Statement}

The data that support the findings of this study are available from the
corresponding author upon reasonable request.

\section*{AUTHOR DECLARATIONS}
\subsection*{Conflict of Interest}
The author have no conflicts to disclose.
\subsection*{Author Contribution}
N.Z. performed the numerical simulations, did the formal analysis, and wrote the first draft of the manuscript.
All authors conceived the project, analyzed the data, and edited the manuscript.


\appendix

\input{supplement.tex}

\end{document}

%% file: supplement.tex
\onecolumngrid

\section{Direct sampling of nucleation times and critical droplet shapes}
\label{sec:stoch-sim}
\begin{figure}
    \centering
    \includegraphics[width=0.8\textwidth]{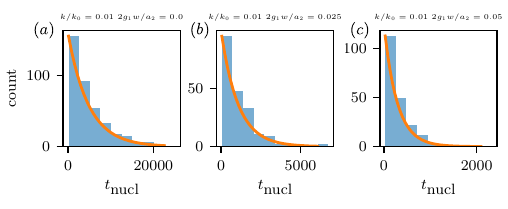}
    \caption{Distribution of nucleation times for different wall affinities $g_1$ and constant non-zero reaction rate $k_0$.}
    \label{fig:si-histogram}
\end{figure}

To extract the mean nucleation time, we used direct sampling of the nucleation events.
To do so, Eq.~\ref{eq:cahn-hilliard-source} in the main text was solved using py-pde \cite{Zwicker_2020}.
The simulation was stopped when a droplet was detected, and the concentration field of the critical droplet was saved. 
We used the mean concentration inside the largest cluster as a nucleation coordinate. 
During nucleation this quantity shows a steep increase which marks the nucleation event.
The nucleation time is not sensitive on the threshold of the order parameter since there is a clear time scale separation between the formation
of a critical nucleus and the growth of the droplet.
The statistics of the nucleation time was extracted over multiple simulations and the time scale of the exponential probability distribution (see Fig. \ref{fig:si-histogram})  function was evaluated.
To get the ensemble of critical nuclei, the contours of the critical droplets were extracted from the critical phase fields using a marching squares algorithm \cite{scikit-image}.  

\section{Constrained optimizations to uncover minimal energy path}
\label{sec:constr-opt}
The minimal energy path comprises a sequence of concentration profiles that connects the homogeneous state to a stationary droplet.
We determine the path using a reaction coordinate $\mathcal{X}$, which measures the mass concentrated in the nucleus, 
\begin{align}
	\mathcal{X}[c] = \int \frac{1}{2}\left[1 + \tanh\left(\Gamma c-\frac{\Gamma}{2\nu}\right)\right] \mathrm{d}V
	\;,
\end{align}
where $\Gamma=10$.
We determine the minimal energy path by minimizing the free energy $\tilde F$ (given in Eq. 9 of the main text) with constrained values of~$\mathcal{X}$.
We impose a value~$\mathcal{X}_0$ of the reaction coordinate using a Lagrange multiplier~$\lambda$ and minimize
\begin{align}
    \hat F[c, \lambda] = \tilde{F}[c] - \lambda (\mathcal{X}[c] - \mathcal{X}_0)
\end{align}
by evolving the partial differential equations
\begin{subequations}
\begin{align}
    \partial_t c &= \Lambda_D \nabla^2\frac{\delta \hat F}{\delta c}
    \;,
\\
    \partial_t \lambda &= -\Lambda_L \frac{\delta \hat F}{\delta \lambda}
    \;,
\end{align}
\end{subequations}
which corresponds to conserved and non-conserved dynamics with mobilities $\Lambda_D$ and $\Lambda_L$, respectively.
We use $\Lambda_D=1$ and $\Lambda_L=100$, which proved a good compromise between speed and stability.
Using this procedure, the profiles~$c(\vect{r})$ that minimize $\hat F$ for each value $\mathcal{X}_0$ of the constraint comprise the minimal free energy path.
The profile with the largest energy~$\tilde F$ corresponds to the saddle point and thus the critical nucleus.

\section{Curvature calculation}
\label{sec:calc-curv}
The curvature $\mathcal{K}$ and the slope $s$ of an implicit curve defined by $c(x_I, y_I) = 0.5$ can be calculated as \cite{Goldman_2005}
\begin{align}
    \mathcal{K} &= \frac{-(\partial_y c)^2 \partial_x^2 c + 2 \partial_x c \partial_y c \partial_{xy} c-(\partial_x c)^2\partial_{y}^2 c}{((\partial_x c)^2 + (\partial_y c)^2)^{(3/2)}}\;,\\
    s &= - \frac{\partial_x c}{\partial_y c}\;,
  \label{eq:slope-imp}
\end{align}
evaluated at $\{x_I, y_I\}$. 
Since the numerical fields are discrete, we use linear interpolation to evaluate the curvature and the slope on the interface line. 

\section{Coupling of chemical reactions and wall affinities are crucial to explain the energy barriers}
\label{sec:coupling}

\subsection{Passive energy barrier}
The passive energy of a sessile droplet can be written as 
\begin{align}
    F = -\Delta f V + \gamma_\mathrm{ds} A_\mathrm{ds} + \gamma_\mathrm{wd} A_\mathrm{wd} + \gamma_\mathrm{ws}(A_\mathrm{w}-A_\mathrm{wd})
    \;,
  \label{eq:pass_het_energy}
\end{align}
where $\Delta f = f(c_0)-f(c_\mathrm{in}^\mathrm{eq})+\partial_c f(c_0)(c_\mathrm{in}^\mathrm{eq}-c_0)$, $A_\mathrm{ds}$ is the droplet solvent interface, $A_\mathrm{wd}$ is the wall droplet interface and $A_\mathrm{w}$ is the total area of the wall. 
Since $\gamma_\mathrm{ds}\cos(\vartheta) = \gamma_\mathrm{ws}-\gamma_\mathrm{wd}$, we can rewrite this expression as 
\begin{align}
    F = -\Delta f V + \gamma_\mathrm{ds} \left(A_\mathrm{ds} - A_\mathrm{wd}\cos(\vartheta)\right) + \gamma_\mathrm{ws}A_\mathrm{w}\;.
\end{align}
The last term only gives a constant offset and will be dropped in the following. For a  spherical cap geometry in two dimensions, we obtain
\begin{align}
    F=\left(\frac12 \Delta f R^2+\gamma_\mathrm{ds}R \right)(2\vartheta - \sin(2\vartheta))
    \;,
\end{align}
and a critical droplet radius of $R_\mathrm{crit} = \gamma_\mathrm{ds}/\Delta f$. The resulting energy barrier is then given by
\begin{align}
    \Delta F = \frac{\gamma_\mathrm{ds}}{2\Delta f}(2\vartheta-\sin(2\vartheta))
    \;.
\end{align}
The wetting angle $\vartheta$ can also be expressed as a function of our microscopic parameters, $\vartheta = \arccos\left(\frac{3g_1}{a_2}\sqrt{\frac{a_4}{2\kappa}}\right)$.

\subsection{Augmented energy barrier by chemical reactions}
The simplest assumption for how chemical reactions and wall affinities affect the energy barrier between homogeneous and droplet state is a simple superposition of the two. We consequently propose
\begin{align}
    \Delta F = \frac{\gamma_\mathrm{ds}^2}{2 \Delta f} (2\vartheta-\sin(2\vartheta)) + mk,
  \label{eq:energy_coupling_naive}
\end{align}
where we for simplicity assumed a linear relation for the reactive energy barrier, $\Delta F_\mathrm{react.}=mk$, in line with our previous results \cite{Ziethen_Kirschbaum_Zwicker_2023}.
We now want to look at contour lines of the energy barrier (where chemical reactions and wall affinity compensate each other) and obtain
\begin{align}
    &k_\mathrm{contour} =\frac{1}{m} \left[ \mathrm{const} - \frac{\gamma^2}{ \Delta f} \left(2\vartheta -\sin(2\vartheta)\right)\right]
    \;.
\end{align}
The second derivative is given by
\begin{align}
    \frac{\partial^2 k_\mathrm{contour}}{\partial g_1^2} = -\frac{2 d^3 g_1 \gamma_\mathrm{ds}^2}{\Delta f \sqrt{1-d^2g_1^2}}<0
    \;,
\end{align}
where $d=\frac{3}{a_2}\sqrt{\frac{a_4}{2\kappa}}$.
Consequently, the contour lines are concave functions of $g_1$, whereas the contours in the data are slightly convex. 

\subsection{Additional coupling}
\begin{figure}
    \centering
    \includegraphics[width=0.8\textwidth]{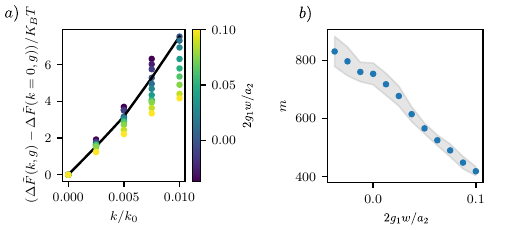}
    \caption{(a) Relative increase of the energy barrier $\Delta \tilde{F}$ with the reaction rate $k$ for different wall affinities $g_1$.
    (b) Fitted slope of the data in panel a as a function of the wall affinity $g_1$.}
    \label{fig:rel_energy_barriers}
\end{figure}
We measured the slope of the energy barrier $m$ for different values of $g_1$ and found a linear decrease (see \figref{fig:rel_energy_barriers}). To capture this observation we add the lowest order coupling between reactions and wall affinities,
\begin{align}
    \Delta F = \Delta F_\mathrm{wall}(g_1) + mk + hkg_1
    \;,
\end{align}
and we obtain for the contour line
\begin{align}
    k_\mathrm{contour} = \frac{1}{m+h g_1} \left[ \mathrm{const} - \Delta F_\mathrm{wall}(g_1)\right]
    \;.
\end{align}
The second derivative in the limit of small wall affinities is then given by
\begin{align}
    \frac{\partial^2 k_\mathrm{contour}}{\partial g_1^2} = \frac{2((\mathrm{const}-\frac{\pi\gamma_\mathrm{ds}^2}{\Delta f})+\frac{\gamma_\mathrm{ds}^2}{\Delta f}d g_1)h^2 - \frac{8\gamma_\mathrm{ds}^2 d h}{\Delta f}(g_1 h +m)}{(g_1 h +m)^3}
    \;,
\end{align}
which for typical parameters is positive and thus shows concave behavior in agreement with the data for $h<0$.

\subsection{Effect of spherical cap shape on reactive energy}
\begin{figure}
    \centering
    \includegraphics[width=0.7\textwidth]{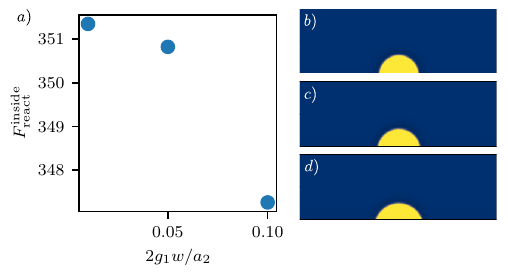}
    \caption{(a) Reactive energy inside a droplet for different wall affinities. (b-d) Corresponding snapshots of droplets.}
    \label{fig:si-coupling-3}
\end{figure}

To test the effect of different contact angles~$\vartheta$ on the reactive energy, we initialized droplets for different wall affinities $g_1$. We then evaluated the reactive energy by integrating Eq.~\ref{eq:energy_react}, neglecting the contribution from the bulk phase. The resulting reactive energy as a function of the wall affinity is depicted in \figref{fig:si-coupling-3}.
\label{sec:react-energy-inside}

\section{Series expansion of reaction-diffusion fluxes}
\label{sec:eff-drop}
\begin{figure}
    \centering
    \includegraphics[width=0.4\textwidth]{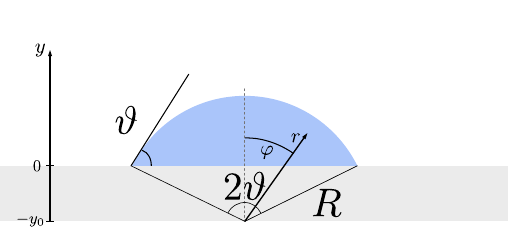}
    \caption{Geometry of a sessile droplet with attractive wall droplet interaction.}
    \label{fig:si-gem}
\end{figure}

In the following, we want to approximate Eq.~\ref{eq:cahn-hilliard-source} of the main text by two reaction-diffusion equations.
We approximate the state of the system by a concentration field inside ($\cIn$) and outside ($\cOut$) the droplet, which are connected by a thin interface.
Since $\cIn$ and $\cOut$ show little variation in their respective phase, we can expand the chemical potential around the respective equilibrium concentration, $\mu(c_i)\approx \partial_c \mu|_{c=c_i^{(0)}}(c-c_i^{(0)})$ \cite{Weber_Zwicker_Juelicher_Lee_2019}.
Inserting this expression into Eq.~\ref{eq:cahn-hilliard-source} of the main text, we obtain the reaction-diffusion equation 
\begin{align}
    \partial_t c_i = D\nabla^2c_i - k(c_i-c_0) \;, \qquad i=\mathrm{in}/\mathrm{out}\;.
  \label{eq:si-react-diff}
\end{align}
The diffusivity $D=\mobD f''(c_\mathrm{in}^{(0)})=\mobD f''(c_\mathrm{out}^{(0)})$ is the same in both phases in our model.
Note that we also neglected the fourth order derivative since we assume low concentration variations.
The geometry of the problem is depicted in \figref{fig:si-gem}. The expected droplet shape follows a circular segment (corresponding to the spherical cap in 3D) with an opening angle of $2\vartheta$, where $\vartheta$ is the contact angle.
The distance between the origin of the circular segment and the droplet wall interface is given by $y_0=R\cos(\vartheta)$. 
Since we assume fast relaxation of the concentration fields inside each phase, we seek a steady state solution of \Eqref{eq:si-react-diff}.
For the steady state equation ($\partial_t c=0$), it is instructive to define the reaction-diffusion length scale $L=\sqrt{D/k}$.
The solution for the concentration field in the given polar geometry is given by \cite{Polyanin_2002}
\begin{align}
    c_i(r, {\varphi})=
   c_0+\sum_{n=0}^{\infty}\left[A_n^i I_{\lambda_n}\left(\frac{r}{L}\right)+B_n^i K_{\lambda_n}\left(\frac{r}{L}\right)\right] 
   \cos (\lambda_n\bigl({\varphi}-\vartheta)\bigr) \;, \qquad i=\mathrm{in}/\mathrm{out}
    \;.
  \label{eq:si-hemholz-sol}
\end{align}
Here, $I_{\lambda_n}(z)$ and $K_{\lambda_n}(z)$ are the modified Bessel functions of first and second kind, and $\lambda_n$ is the separation variable which has to fulfill $\lambda_n=n\pi/\vartheta$ to be consistent with the boundary conditions.
We next will determine the coefficients $A_n^i$ and $B_n^i$ from the boundary conditions.

\subsection{Concentration field inside the droplet}
The relevant boundary conditions are
\begin{align}
    \left. \frac{\partial c}{\partial y}\right |_{y=0} &= 0 &
    c(r=R)&=\cIn^{(0)} + \beta_\mathrm{in}\gamma_\mathrm{ds}/R\;.
  \label{eq:bc0}
\end{align}
Translating the first boundary condition to polar coordinates gives
\begin{align}
    \left. \frac{\partial c(r(x,y), \vpp(x,y)}{\partial y}\right|_{0}
    	= \frac{\partial c}{\partial r}\frac{\partial r}{\partial y} + \frac{\partial c}{\partial \varphi}\frac{\partial \varphi}{\partial y}
	= \left[\frac{\partial c}{\partial r} \cos(\vpp) - \frac{\partial c}{\partial \vpp} \frac{\sin(\vpp)}{r}\right]_{r_0(\vpp)=R\frac{\cos(\vartheta)}{\cos(\vpp)}}
	=0\;.
  \label{eq:bc_polar}
\end{align}
We rewrite the second boundary condition as 
\begin{align}
    \sum_n A_n d_n(\vpp) + B_n e_n(\vpp) = \cIn^{(0)} + \beta_\mathrm{in}\gamma_\mathrm{ds}/R\;,
\end{align}
with
\begin{align}
    d_n(\vpp) &= I_{\lambda_n}(R/L)\cos({\lambda_n}\vpp) &
    e_n(\vpp) &= K_{\lambda_n}(R/L)\cos({\lambda_n}\vpp)
    \;.
\end{align}
Next, we define the projection operator 
\begin{align}
    P[f]\equiv \int_{-\vartheta}^{\vartheta} f(\vpp)h_m(\vpp) \text{d}\vpp
\end{align}
for a well behaving function $h_m(\vpp)$, which will be defined explicitly later.
Applying the projection operator to the boundary condition leads to
\begin{multline}
    \sum_n \left[A^n \left(I_{\lambda_n}(R/L)\int_{-\vartheta}^{\vartheta}  \cos({\lambda_n}\vpp)h_m(\vpp) \text{d}\vpp\right)  
    + B^n \left(K_{\lambda_n}(R/L)\int_{-\vartheta}^{\vartheta}  \cos({\lambda_n}\vpp)h_m(\vpp) \text{d}\vpp\right) \right]
   \\
    = \left(\cIn^{(0)}+\frac{\beta_\mathrm{in}\gamma_\mathrm{ds}}{R}-c_0\right)\int_{-\vartheta}^{\vartheta}  h_m(\vpp) \text{d}\vpp
    \;.
\end{multline}
We rewrite this expression as
\begin{align}
  \sum_n (C_{mn} A_n  + D_{mn}B_n ) = b_m\;
  \Leftrightarrow\;\mathbf{C} A + \mathbf{D}B = b
\end{align}
with coefficients
\begin{subequations}
\begin{align}
  C_{mn}&=\int_{-\vartheta}^{\vartheta} d_n(\vpp) h_m(\vpp) \text{d}\vpp\;,\\
  D_{mn}&=\int_{-\vartheta}^{\vartheta} e_n(\vpp) h_m(\vpp) \text{d}\vpp\;,\\
  b_m &= \left(\cIn^{(0)}+\frac{\beta_\mathrm{in}\gamma_\mathrm{ds}}{R}-c_0\right)\int_{-\vartheta}^{\vartheta}  h_m(\vpp) \text{d}\vpp\;.
\end{align}
\end{subequations}
Consequently, we recast the boundary condition into a system of linear equations. The number of equations is in principle infinite, but we can hope to obtain an accurate approximation by looking at finite number $m$ of modes.
We want to treat the droplet-wall boundary condition in a similar way.
We can insert the solution for the concentration field into the boundary condition and obtain
\begin{multline}
    \sum_n \left[ A^n \frac{I_{\lambda_n-1}\left(\frac{R\cos(\vartheta)}{L\cos(\vpp)}\right) +  I_{\lambda_n+1}\left(\frac{R\cos(\vartheta)}{L\cos(\vpp)}\right)}{2L}-  B^n \frac{K_{\lambda_n-1}\left(\frac{R\cos(\vartheta)}{L\cos(\vpp)}\right) +  K_{\lambda_n+1}\left(\frac{R\cos(\vartheta)}{L\cos(\vpp)}\right)}{2L}\right]\cos(\lambda_n \vpp) \\
    + \sum_n \left[A^n I_{\lambda_n} \left(\frac{R\cos(\vartheta)}{L\cos(\vpp)}\right) + B^n K_{\lambda_n}\left(\frac{R\cos(\vartheta)}{L\cos(\vpp)}\right)\right]\frac{\sin(\lambda_n \vpp)\sin(\vpp)}{R\cos(\vartheta)}=0
    \;,
\end{multline}
which we rewrite as 
\begin{align}
    \sum_n A_n f_n(\vpp) + B_n g_n(\vpp) = 0 \;,
\end{align}
with
\begin{subequations}
\begin{align}
    f_n(\vpp) &= \frac{I_{\lambda_n-1}\left(\frac{R\cos(\vartheta)}{L\cos(\vpp)}\right) +  I_{\lambda_n+1}\left(\frac{R\cos(\vartheta)}{L\cos(\vpp)}\right)}{2L}\cos(\lambda_n \vpp) + I_{\lambda_n}\left(\frac{R\cos(\vartheta)}{L\cos(\vpp)}\right)\frac{\sin(\lambda_n \vpp)\sin(\vpp)}{R\cos(\vartheta)}\;,\\
    g_n(\vpp) &= \frac{-K_{\lambda_n-1}\left(\frac{R\cos(\vartheta)}{L\cos(\vpp)}\right) -  K_{\lambda_n+1}\left(\frac{R\cos(\vartheta)}{L\cos(\vpp)}\right)}{2L}\cos(\lambda_n \vpp) + K_{\lambda_n}\left(\frac{R\cos(\vartheta)}{L\cos(\vpp)}\right)\frac{\sin(\lambda_n \vpp)\sin(\vpp)}{R\cos(\vartheta)}
    \;.
\end{align}
\end{subequations}
After applying the projection operator $P$, we then obtain another system of linear equations
\begin{align}
    \sum_n E_{mn}A_n  + F_{mn}B_n  = 0\;
    \Leftrightarrow \mathbf{E}A + \mathbf{F} B = 0\;,
\end{align}
where the coefficients are given by 
\begin{subequations}
\begin{align}
  E_{mn} &=\int_{-\vartheta}^{\vartheta} f_n(\vpp) h_m(\vpp) \text{d}\vpp\;,\\
  F_{mn} &= \int_{-\vartheta}^{\vartheta} g_n(\vpp) h_m(\vpp) \text{d}\vpp\;.
\end{align}
\end{subequations}
Finally, the only task left to do is to solve the two matrix equations
\begin{subequations}
\begin{align}
  \mathbf{C} A + \mathbf{D}B = b \;,\\
\mathbf{E}A + \mathbf{F} B = 0 \;.
\end{align}
\end{subequations}
We can combine both equations to 
\begin{align}
    {\scriptstyle
\begin{pmatrix}
    \mathbf{C} & \mathbf{D} \\
    \mathbf{E}  & \mathbf{F} 
\end{pmatrix}
\begin{pmatrix}
    A \\  B
\end{pmatrix}
=
\begin{pmatrix}
    b \\ 0
\end{pmatrix}\;,
}
\end{align}
and invert the full matrix numerically. We choose $h_m(\vpp)=\cos(\pi m\vpp/\vartheta)$ in the projection operator and $m\in [0,3]$.

\subsection{Concentration field outside the droplet}
\label{sec:attr-wall-out}
For the concentration profile outside we have in principle three boundary conditions
\begin{align}
    c(R) &= \cOut^{(0)} + \beta_\mathrm{out} \gamma_\mathrm{ds}/R &
    \left.\frac{\partial c}{\partial z}\right|_{z=z_0} &= 0 &
    \lim_{r \rightarrow \infty}\frac{\partial c}{\partial r}&= 0\;. 
\end{align}
The projection method we used before is not well suited for this situation since the first and the last boundary condition are defined over $\vpp \in [-\vartheta, \vartheta]$, whereas the second one is defined over $\vpp \in [-\pi/2, -\vartheta]\cup[\vartheta, \pi/2]$. We can define the operators over the specified range of $\vpp$ and get three equations for two unknowns, which gives an overdetermined system.
One way around this is to consider a domain given by the opening angle $2\vartheta$. If we use the same quantization of the separation variable $\lambda=\pi n/\vartheta$, we can solve the problem with the two boundary conditions which are defined over the opening angle.
Using the same operator as defined for the inside concentration we get the two equations
\begin{subequations}
\begin{align}
    \mathbf{C} A + \mathbf{D}B &= \mathbf{b}_{out}\;,\\
\mathbf{G}A + \mathbf{H} B &= 0\;,
\end{align}
\end{subequations}
where
\begin{subequations}
\begin{align}
    G_{mn} &= \int_{-\vartheta}^{\vartheta} k_n(\vpp)h_m(\vpp) \text{d}\vpp\;,\\
    H_{mn} &= \int_{-\vartheta}^{\vartheta} l_n(\vpp)h_m(\vpp) \text{d}\vpp\;,\\
    b_m^\mathrm{out} &= \left(\cOut^{(0)}+\frac{\beta_\mathrm{in}\gamma_\mathrm{ds}}{R}-c_0\right)\int_{-\vartheta}^{\vartheta}  h_m(\vpp) \text{d}\vpp\;,\\
\end{align}
\end{subequations}
with
\begin{subequations}
\begin{align}
    k_n(\vpp) &= \frac{I_{\lambda_n-1}\left(\frac{R_\infty}{L}\right) +  I_{\lambda_n+1}\left(\frac{R_\infty}{L}\right)}{2L}\cos(\lambda_n \vpp)\;,\\
    l_n(\vpp) &= \frac{-K_{\lambda_n-1}\left(\frac{R_\infty}{L}\right) -  K_{\lambda_n+1}\left(\frac{R_\infty}{L}\right)}{2L}\cos(\lambda_n \vpp)\;.
\end{align}
\end{subequations}
We can solve for the coefficients as before by inverting the combined matrix. Note that $\mathbf{b}_\mathrm{out}$ is defined as before but with a different constant value.
\begin{align}
	{\scriptstyle
		\begin{pmatrix}
			\mathbf{C} & \mathbf{D} \\
			\mathbf{G}  & \mathbf{H} 
		\end{pmatrix}
		\begin{pmatrix}
			A \\  B
		\end{pmatrix}
		=
		\begin{pmatrix}
			b_\mathrm{out} \\ 0
		\end{pmatrix}\;.
	}
\end{align}

\subsection{Special case of a neutral wall}
In the case of a neutral wall the opening angle is $2\vartheta=\pi$, and we obtain a slightly different matrix equation for the coefficient, which needs extra treatment. 

\subsubsection{Concentration field inside the droplet}
The solution to the Helmholtz equation is still given by the concentration field
\begin{align}
  c(r, \vpp) = c_0 + \sum_{n=0}^\infty (A^n I_\lambda(r/L) + B^n K_\lambda(r/L))\cos(\lambda \vpp)
  \;.
\end{align}
For the neutral wall, we obtain the boundary conditions for the concentration profile inside the droplet as
\begin{align}
    \left.\frac{\partial c}{\partial r}\right|_{r=0}&=0
    &
    \left.\frac{\partial c}{\partial \vpp}\right|_{\vpp=\pm\pi/2}&=0
    \;.
\end{align}
This allows us to eliminate $B^n=0$. The second boundary condition is already fulfilled by the choice of the separation variable $\lambda_n=n\pi/\vartheta$. We are left with the boundary condition
\begin{align}
    {c}(R)&=\cIn^{(0)}-c_0+\beta\mathrm{in} \gamma_\mathrm{ds}/R\notag\\
          &=\sum_n A_n I_{\lambda_n}(R/L)\cos(\lambda_n\vpp) = \cIn^{(0)}-c_0+\beta_\mathrm{in} \gamma_\mathrm{ds}/R
          \;.
\end{align}
We can then apply the projection $P$ operator and get
\begin{align}
    \begin{split}
    \sum_n A_n I_{\lambda_n}(R/L)\int_{-\pi/2}^{\pi/2} \cos(\lambda_n\vpp) h_m(\vpp) \text{d}\vpp
    = (-c_0+\beta \gamma/R)\int_{-\pi/2}^{\pi/2} h_m(\vpp) \text{d}\vpp 
\end{split}
\end{align}
The corresponding matrix equation can be written as 
\begin{align}
    \sum_n \bar{C}_{mn} A_n  = b_m\; \Leftrightarrow \;
    \mathbf{\bar{C}}A=b
\end{align}
with coefficients 
\begin{subequations}
\begin{align}
    \bar{C}_{mn}&=I_{\lambda_n}(R/L)\int_{-\pi/2}^{\pi/2} \cos(\lambda_n\vpp) h_m(\vpp) \text{d}\vpp \;,\\
    b_{m} &= (\cIn^{(0)}-c_0+\beta \gamma/R)\int_{-\pi/2}^{\pi/2} h_m(\vpp) \text{d}\vpp \;. 
\end{align}
\end{subequations}

\subsubsection{Concentration field outside the droplet}
The concentration outside $\cOut$ can now be solved for the full domain. Fortunately, the equations remain the same as in the case of an attractive wall, see Sec.~\ref{sec:attr-wall-out}.

All matrix equations where then solved using numpy for the first four modes \cite{harris2020array}.

\section{Reactions deform sessile 3D droplets}
\begin{figure}
    \centering
    \includegraphics[width=0.8\textwidth]{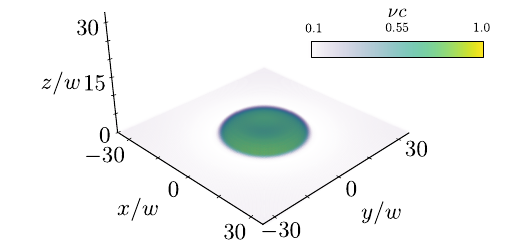}
    \caption{Concentration profile $c(x, y, z)$ of a stationary 3D droplet for reaction rate $k/k_0=0.0053$ and wall affinity $g_1=0.05a_2/w$. Simulations were performed in Cartesian coordinates using $L_\mathrm{x}=L_\mathrm{y}=64w$, $L_\mathrm{z}=32w$ with a discretization of $0.5w$.}
    \label{fig:si-3d}
\end{figure}

To test whether the observed droplet deformations also persist in 3D, we first performed simulations in Cartesian coordinates. 
\figref{fig:si-3d} shows that we indeed find deformed, flattened droplet shapes as expected from our analysis in 2D.
Since the simulations exhibit a cylindrical symmetry, we used cylindrical coordinates for the simulations shown in the main text to minimize the computational cost. 

\section{Electrostatic analogy explains effects of driven reactions}
\label{sec:scaling}

In this section, we investigate the equilibrium surrogate model given by \Eqref{eq:tot_en_func} in the main text in detail, allowing us to interpret the reactions as long-ranged electrostatic interactions.
By adopting this energy perspective, we can assess the conditions under which droplet deformation and droplet splitting might be favorable.

\subsection{Geometry of the problem}

\begin{figure}
    \centering
    \includegraphics[width=0.9\textwidth]{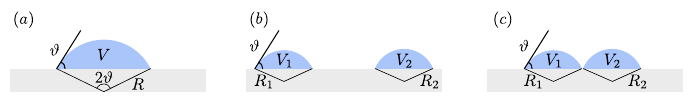}
    \caption{Geometry of the one droplet state (a) compared to the two separated droplets (b) and two connected droplets.}
    \label{fig:geom}
\end{figure}

We want to compare the energy of the three scenarios depicted in Fig.~\ref{fig:geom}:
The first one consist of a single droplet (I), the second one consists of two smaller separated droplets (II), and the third one consists of two smaller adjacent droplets (II-c).
We assume that the contact angle $\vartheta$ is constant and only depends on $g_1$ (as defined in the main text).
Since we want to discuss the 2D and 3D case alongside, we introduce $V_\mathrm{2D}$ for the area of a circular segment and $V_\mathrm{3D}$ as the volume of a spherical cap.
For a circular segment/spherical cap, we can calculate the volume $V$, the arc length $a$ (surface of sphere in 3D), and the chord length $c$ as (surface of droplet-wall interface),
\begin{subequations}
\begin{align}
    V_\mathrm{2D} &= \frac{R^2}{2}(2\vartheta - \sin(2\vartheta))\;,&
    a_\mathrm{2D} &= 2\vartheta R\;,&
    c_\mathrm{2D} &= 2R\sin(\vartheta) \;.
\\
    V_\mathrm{3D} &= \frac{\pi}{3}R^3(2+\cos(\vartheta))(1-\cos(\vartheta))^2\;,&
    a_\mathrm{3D}&= 2\pi R^2 (1-\cos(\vartheta))\;,&
    c_\mathrm{3D} &= \pi \sin^2(\vartheta)R^2 \;.
\end{align}
\end{subequations}
Since we assume constant total volume, we can relate the radii of the two smaller droplets ($R_1$ and $R_2$) to the radius of the single droplet by $R_{1}=R_2=R/\sqrt{2}$ in 2D and by $R_1=R_2=R/\sqrt[3]{2}$ in 3D. 
The distance between the centers of the two droplets is given by $l=2R_{1}\sin(\vartheta)$ for the adjacent droplets since we assume that they touch each other. 
For the separated droplets we assume no interaction between them.
We consider a wall of length $L_\mathrm{sys}$.

\subsection{Energy contributions for a single droplet}
We expect that the behavior of the system is governed by three energy contributions: interfacial energy, volume energy, and a contribution that captures the influence of reactions.
The volume energy is given by 
\begin{subequations}
\begin{align}
    E_V^\mathrm{2D}&=-\Delta f V_\mathrm{2D} = -\Delta f \frac{R^2}{2}(2\vartheta - \sin(2\vartheta)) \;,\\
    E_V^\mathrm{3D}&=-\Delta f V_\mathrm{3D} = -\Delta f \frac{\pi}{3}R^3(2+\cos(\vartheta))(1-\cos(\vartheta))^2 \;.
\end{align}
\end{subequations}
The interfacial energy is given by $E_\mathrm{int}=\gamma_\mathrm{ds} a + \gamma_\mathrm{wd} c + \gamma_\mathrm{ws}(L_\mathrm{sys}-c)$.
Since $\gamma_\mathrm{ds}\cos(\vartheta) = \gamma_\mathrm{ws}-\gamma_\mathrm{wd}$, we generally obtain $E_\mathrm{int} = \gamma_\mathrm{ds}(a-c \cos(\vartheta)) + \gamma_\mathrm{ws}L_\mathrm{sys}$, which for the concrete systems read
\begin{subequations}
\begin{align}
    E^\mathrm{2D}_\mathrm{int} &= \gamma_\mathrm{ds}R(2\vartheta-\sin(2\vartheta)) + \gamma_\mathrm{ws}L_\mathrm{sys} \;,\\
    E^\mathrm{3D}_\mathrm{int} &= \gamma_\mathrm{ds}4\pi R^2\sin(\vartheta/2)^4 + \gamma_\mathrm{ws}L_\mathrm{sys} \;.
\end{align}
\end{subequations}
We can also express the interfacial energy as a function of the volume by substituting $R = \sqrt{2V^\mathrm{2D}/(2\vartheta - \sin(2\vartheta)}$ and $R= \left(\frac{3V_\mathrm{3D}}{\pi(2+\cos(\vartheta)(1-\cos(\vartheta)^2)}\right)^{(2/3)}$,
\begin{subequations}
\begin{align}
    E^{2D}_\mathrm{int} &= \gamma_\mathrm{ds}\sqrt{2V_\mathrm{2D}(2\vartheta-\sin(2\vartheta))} \;, \\
    E^{3D}_\mathrm{int} &=4\pi \left(\frac{3V_\mathrm{3D}}{\pi(2+\cos(\vartheta))(1-\cos(\vartheta)^2)}\right)^{(2/3)}\sin^4(\vartheta/2) \;.
\end{align}
\end{subequations}
The reactive energy can be determined from the electrostatic analogy, but we do not have a precise expression.
To derive a scaling relation, we use the electrostatic energy of a droplet on a neutral wall, which serves as an upper bound for the energy of a spherical cap.

\subsubsection{Energy of single chemically active droplet in dilute phase in two-dimensional system}
As established in \ref{sec:surr-model}, we can map the chemically active system onto an electrostatic system with the energy
\begin{align}
  F_\mathrm{react} = \frac{k}{2\Lambda_\mathrm{d}}\int (c-c_0)\Psi \text{d}V\;,
  \label{eq:elect_en_si}
\end{align}
with 
\begin{align}  
\nabla^2 \Psi = -(c-c_0)\;.
\label{eq:poisson}
\end{align}
We want to solve this equation and consequently need an expression for the concentration field, which we can approximate as solution to reaction-diffusion equations 
\begin{align}
    \partial_t c_i = D \nabla^2 c_i -k(c-c_0)\;,~ i\in \{\text{inside}, \text{outside}\}
\end{align}
in stationary state, i.e., we assume \emph{quasi-stationarity}.
The solutions, with vanishing derivative at the origin and infinity and $\cIn(R)= \cIn^{(0)}$, $\cOut(R)= \cOut^{(0)}$, are given by
\begin{subequations}
\begin{align}
  c_\mathrm{in} = c_0 + (c_\mathrm{in}^\mathrm{(0)}-c_0)\frac{I_0(r/L)}{I_0(R/L)}\;,\\
  c_\mathrm{out} = c_0 + (c_\mathrm{out}^\mathrm{(0)}-c_0)\frac{K_0(r/L)}{K_0(R/L)}
  \;,
\end{align}
\end{subequations}
where $L=\sqrt{D/k}$ is the reaction-diffusion length scale as defined  in the main text.
These concentration profiles reproduce the main features of the solution to the full phase field equations. However, they do not conserve mass in the current form, which is easily seen by calculating the integrated reactive current inside and outside,
\begin{subequations}
\begin{align}
  S_\mathrm{in} &= -\frac{k 2\pi (c_\mathrm{in}^\mathrm{(0)}-c_0)}{I_0(R/L)}\int_0^R {I_0(r/L)}r \text{d}r=-\frac{k 2\pi (c_\mathrm{in}^\mathrm{(0)}-c_0)}{I_0(R/L)}L R I_1(\frac{R}{L})\;,\\
S_\mathrm{out} &= -\frac{k 2\pi (c_\mathrm{out}^\mathrm{(0)}-c_0)}{K_0(R/L)}\int_R^\infty {K_0(r/L)}r \text{d}r = -\frac{k 2\pi (c_\mathrm{out}^\mathrm{(0)}-c_0)}{K_0(R/L)}L R K_1(R/L)
	\;.
\end{align}
\end{subequations}
This breaking of mass conservation is inconsistent with the initial equations \Eqref{eq:cahn-hilliard-source} and \Eqref{eq:reaction_flux_lin} in the main text, which conserve the overall mass if the average density in the system is $c_0$.
In contrast, the thin-interface model with quasi-stationary that we here use for simplicity breaks this mass conservation.
Since this is a crucial feature of our system, we correct the mass by demanding a modified equilibrium concentration outside the droplet,
\begin{align}
  c_\mathrm{out}^\mathrm{(0), mod} = -(c_\mathrm{in}^\mathrm{(0)}-c_0)\frac{I_1(R/L)K_0(R/L)}{I_0(R/L)K_1(R/L)} + c_0
  \;,
\end{align}
which we use in the following. We can then solve the energy integral
\begin{align}
  F_\mathrm{react} = \frac{k}{2\Lambda_\mathrm{d}}\int (c-c_0)\Psi \text{d}V\;,
\end{align}
by rewriting it as 
\begin{align}
      F_\mathrm{react} = -\frac{k}{2\Lambda_\mathrm{d}}\int (\nabla^2 \Psi) \Psi \text{d}V\;.
  \end{align}
The function $c-c_0$ is not continuous, so we integrate piecewise by parts
\begin{align}
    F_\mathrm{react} &= -\frac{k}{2\Lambda_\mathrm{d}}\left(\int_{V_\mathrm{in}} (\nabla^2 \Psi) \Psi \text{d}V + \int_{V_\mathrm{out}} (\nabla^2 \Psi) \Psi \text{d}V\right)\;,
\notag\\
      &= -\frac{k}{2\Lambda_\mathrm{d}} \left(\int_{\partial V_\mathrm{in}} \Psi \nabla \Psi \mathbf{n} \text{d}A - \int_{V_\mathrm{in}} \nabla \Psi \nabla \Psi \text{d}V + \int_{\partial V_\mathrm{out}} \Psi \nabla \Psi \mathbf{n} \text{d}A - \int_{V_\mathrm{out}} \nabla \Psi \nabla \Psi \text{d}V \right)\;.
\end{align}
Using polar coordinates and the fact that we have Neumann conditions on $\Psi$ at $R=0$ and $R\to\infty$, we find
\begin{align}
     F_\mathrm{react} = -\frac{k}{2\Lambda_\mathrm{d}} \left(2\pi R (\Psi_\mathrm{in} \partial_r \Psi_\mathrm{in}-\Psi_\mathrm{in} \partial_r \Psi_\mathrm{in})  - \int_{V_\mathrm{in}} (\partial_r \Psi_\mathrm{in})^2\text{d}V - \int_{V_\mathrm{out}} (\partial_r \Psi_\mathrm{out})^2\text{d}V \right)
     \;.
  \label{eq:partial_int_boundary}
\end{align}
We next define the potential $\Omega=\partial_r\Psi$, plug it into the Poisson equation given by \Eqref{eq:poisson},
and solve for $\Omega$,
\begin{subequations}
\begin{align}
  \partial_r \Omega_\mathrm{in} + \frac{\Omega_\mathrm{in}}{r} &= -(c_\mathrm{in}^\mathrm{(0)}-c_0)\frac{I_0(r/L)}{I_0(R/L)}\;,\\
  \Omega_\mathrm{in} &= \frac{d_2}{r}-(c_\mathrm{in}^\mathrm{(0)}-c_0)\frac{I_1(r/L)L}{ I_0(R/L)}
\;,
\end{align}
\end{subequations}
where $d_2$ is an integration constant. The constant $d_2$ needs to be zero since the field needs to be divergence-free at $r=0$. 
Similarly, we obtain 
\begin{align}
    \Omega_\mathrm{out} = -\frac{L (c_0-{\cOut})
   K_1\left(\frac{r}{L}\right)}{K_0\left(\frac{R}{L}\right)}\;.
\end{align}
We immediately see that with the modified outside concentration we have $\Omega_\mathrm{in}(R)=\Omega_\mathrm{out}(R)$ and the boundary term in \Eqref{eq:partial_int_boundary} vanish. 
Importantly, the boundary terms do not vanish by themselves but only after integration over full space.
We thus find
\begin{align}
    F_\mathrm{react} = -\frac{k}{2\Lambda_\mathrm{d}} \left( - \int_{V_\mathrm{in}} (\partial_r \Psi_\mathrm{in})^2\text{d}V - \int_{V_\mathrm{out}} (\partial_r \Psi_\mathrm{out})^2\text{d}V \right)
    \;.
\end{align}
We then integrate the inside contribution of the energy,
\begin{align}
    F_\mathrm{react}^\mathrm{in} &= \frac{2\pi k}{2\Lambda_\mathrm{d}} \int_0^R  \Omega_\mathrm{in}^2 r \text{d}r = \frac{2\pi k}{2\Lambda_\mathrm{d}} \left(\frac{(\cIn-c_0)L}{I_0(R/L)}\right)^2\frac{1}{2} R \left(-R I_0\left(\frac{R}{L}\right){}^2+2 L I_1\left(\frac{R}{L}\right)
   I_0\left(\frac{R}{L}\right)+R I_1\left(\frac{R}{L}\right){}^2\right)\;, \notag\\
   &\approx \frac{\pi k}{16\Lambda_\mathrm{d}}\left(\cIn-c_0\right)^2R^4\;, 
\end{align}
where the last approximation holds for small droplet radii $R$ and is in general a lower bound to the entire function. 
For the outside contribution we get
\begin{align}
    F_\mathrm{react}^\mathrm{out} &= \frac{2\pi k}{2\Lambda_\mathrm{d}} \left(\frac{(\cOut-c_0)L}{K_0(R/L)}\right)^2\int_R^\infty K_1(r/L)^2 r \text{d}r\;,
\notag\\
    &=\frac{2\pi k}{2\Lambda_\mathrm{d}} \left(\frac{(\cOut-c_0)L}{K_0(R/L)}\right)^2 \left(\frac{1}{2} R \left(R K_0\left(\frac{R}{L}\right){}^2+2 L K_1\left(\frac{R}{L}\right)
    K_0\left(\frac{R}{L}\right)-R K_1\left(\frac{R}{L}\right){}^2\right)\right)\;,
\notag\\
    &=\frac{\pi k}{\Lambda_\mathrm{d}} \left({(\cOut-c_0)L}\right)^2 R^2\left( \frac{1}{2}+ L \frac{K_1\left(\frac{R}{L}\right)}{
K_0\left(\frac{R}{L}\right)R}- \frac{1}{2} \frac{K_1\left(\frac{R}{L}\right){}^2}{K_0\left(\frac{R}{L}\right){}^2}\right)\;.
\end{align}
To determine the scaling behavior of this expression with $R$, we need to plug in the modified outside concentration
\begin{align}
    F_\mathrm{react}^\mathrm{out}&=\frac{\pi k(c_\mathrm{in}^\mathrm{(0)}-c_0)^2L^2}{\Lambda_\mathrm{d}} \left({\frac{I_1(R/L)K_0(R/L)}{I_0(R/L)K_1(R/L)}}\right)^2 R^2\left( \frac{1}{2}+ L \frac{K_1\left(\frac{R}{L}\right)}{
K_0\left(\frac{R}{L}\right)R}- \frac{1}{2} \frac{K_1\left(\frac{R}{L}\right){}^2}{K_0\left(\frac{R}{L}\right){}^2}\right)\;
\notag\\
&\approx\frac{\pi k(c_\mathrm{in}^\mathrm{(0)}-c_0)^2}{\Lambda_\mathrm{d}8}
{\left(2 \log \left(\frac{2 L}{R}\right)-(2\gamma_\mathrm{euler} +1)\right)}R^4 
\end{align}
Consequently the total energy scaling is given by
\begin{align}
F_\mathrm{react}^\mathrm{tot} = \frac{\pi k(c_\mathrm{in}^\mathrm{(0)}-c_0)^2}{8 \Lambda_\mathrm{d}}\left({2 \log \left(\frac{2 L}{R}\right)-(2 \gamma_\mathrm{euler} +\frac{1}{2})}\right)R^4
  \label{eq:energy-total-2d}
\end{align}
where $\gamma_\mathrm{euler}\approx 0.577$ is the Euler number. 
This expression is again valid for small droplet radii but the approximation becomes unphysical as one approaches the reaction-diffusion length scale $L=\sqrt{D/k}$. 
The approximation then becomes negative, whereas the full expression approaches an almost linear positive increase. 

\subsubsection{Energy of single chemically active droplet in dilute phase in three-dimensional system}
We perform the same calculations as above for three dimensions, which will give slight modifications.
We again start from the two reaction-diffusion equations
\begin{align}
\partial_t c_i = D \nabla^2 c_i -k(c-c_0);,~ i\in \{\text{inside}, \text{outside}\}.
\end{align}
We can write these equations in spherical coordinates and then solve for the stationary state
\begin{align}
\partial_r^2 c + \frac{2}{r}\partial_r c -\frac{1}{L^2}(c-c_0)=0\;.
\end{align}
With boundary conditions $\partial r c_\mathrm{in}|_{r=0} = 0$, $c(R)=c_\mathrm{in}$ the solution is given by
\begin{align}
    c_\mathrm{in} = \frac{R ({c_\mathrm{in}^\mathrm{(0)}}-{c_0}) \sinh \left(\frac{r}{L}\right)
   \text{csch}\left(\frac{R}{L}\right)}{r}+{c_0}\;.
\end{align}
For the outside, with boundary conditions $\lim_{r \to \infty} \partial_r c_\mathrm{in}|_{r=0} = 0$ and $c(R)=c_\mathrm{out}^\mathrm{(0)}$, we find 
\begin{align}
c_\mathrm{in} = c_0 + (c_\mathrm{out}^\mathrm{(0)}-c_0)\frac{\exp((-r+R)/L)R}{r}\;.
\end{align}
From the concentration fields, we can next calculate the integrated reactive flux for the inside
\begin{align}
S_\mathrm{in} &= \int_{V_\mathrm{in}} -k(c_\mathrm{in}^\mathrm{(0)}-c_0) \text{d}V\;,
\notag\\
&= -4\pi k R \frac{(c_\mathrm{in}^\mathrm{(0)}-c_0)}{\sinh(R/L)} \int_0^R \frac{\sinh(r/L)}{r} r^2 \text{d}r\;,
\notag\\
&= -4\pi k R \frac{(c_\mathrm{in}^\mathrm{(0)}-c_0)}{\sinh(R/L)} L \left(R \cosh \left(\frac{R}{L}\right)-L \sinh
   \left(\frac{R}{L}\right)\right)\;,
\notag\\
&=-4\pi k R {(c_\mathrm{in}^\mathrm{(0)}-c_0)} L \left(R \coth \left(\frac{R}{L}\right)-L\right)\;,
\end{align}
and the outside
\begin{align}
S_\mathrm{out} &= \int_{V_\mathrm{out}} -k(c_\mathrm{out}-c_0) \text{d}V\;,
\notag\\
&= -4\pi k (c_\mathrm{out}^\mathrm{(0)}-c_0)R \exp(R/L)\int_R^{\infty} \exp(-r/L)r \text{d}r\;,
\notag\\
&=-4\pi k (c_\mathrm{out}^\mathrm{(0)}-c_0)R \exp(R/L)L^2\;.
\end{align}
To balance the two fluxes, which ensures the solvability of the Poisson problem, we obtain a modified outside concentration,
\begin{align}
c_\mathrm{out}^\mathrm{(0), mod} &= \frac{(c_\mathrm{in}^\mathrm{(0)}-c_0)}{\sinh(R/L)}  \left(R \cosh \left(\frac{R}{L}\right)-L \sinh\left(\frac{R}{L}\right)\right)\frac{\exp(-R/L)}{L} + c_0\;,
\notag\\
&= \frac{e^{-\frac{R}{L}} \left(c_{\text{in}}^{\text{(0)}}-c_0\right) \left(L-R
   \coth \left(\frac{R}{L}\right)\right)}{L} +c_0\;.
\end{align}
Next, we need to calculate the potential $\Omega=\partial_r \Psi$.
We get for the inside 
\begin{align}   
\Omega_\mathrm{in} = -\frac{R ({c_\mathrm{in}^\mathrm{(0)}}-{c_0})}{\sinh(R/L)}\frac{L r \cosh \left(\frac{r}{L}\right)-L^2 \sinh
   \left(\frac{r}{L}\right)}{r^2}\;,
\end{align}
and for the outside
\begin{align}
\Omega_\mathrm{out} =  (c_\mathrm{out}^\mathrm{(0)}-c_0){\exp((R)/L)R} \frac{ L e^{-\frac{r}{L}} (L+r)}{r^2}\;.
\end{align}
The inside energy is then given by
\begin{align}
F_\mathrm{react}^\mathrm{in} &= \frac{4\pi k}{\Lambda_\mathrm{d}2} \int_0^R \Omega^2 r^2 \text{d}r\;,
\notag\\
&=\frac{4\pi k}{\Lambda_\mathrm{d}2} \left(\frac{R L (c_\mathrm{in}^\mathrm{(0)}-c_0)}{\sinh(R/L)}\right)^2\int \frac{(r \cosh(r/L)-L\sinh(r/L))^2}{r^2} \text{d}r\;,
\notag\\
&=\frac{\pi k L^2 (c_\mathrm{in}^\mathrm{(0)}-c_0)^2}{2\Lambda_\mathrm{d}} \left(\frac{R }{\sinh(R/L)}\right)^2  \left(-\frac{4 L^2 \sinh ^2\left(\frac{R}{L}\right)}{R}+L \sinh
   \left(\frac{2 R}{L}\right)+2 R\right)\;,
\notag\\
   &\approx \frac{\pi k  (c_\mathrm{in}^\mathrm{(0)}-c_0)^2}{2\Lambda_\mathrm{d}} \frac{4}{45} R^5\;,
\end{align}
where we expanded for small droplet radii in the last line.
The outside energy can be expressed as
\begin{align}
F_\mathrm{react}^\mathrm{out} &= \frac{4\pi k}{\Lambda_\mathrm{d}2} \int_0^R \Omega^2 r^2 \text{d}r\;,
\notag\\
&=\frac{4\pi k}{\Lambda_\mathrm{d}2}\left((c_\mathrm{out}^\mathrm{(0)}-c_0){\exp((R)/L)R L}\right)^2 \int_R^\infty \left(\frac{ e^{-\frac{r}{L}} (L+r)}{r^2}\right)^2r^2 \text{d}r\;,
\notag\\
&= \frac{4\pi k (c_\mathrm{out}^\mathrm{(0)}-c_0)^2 L^2}{\Lambda_\mathrm{d}2}\left({\exp((R)/L)R }\right)^2 \frac{L e^{-\frac{2 R}{L}} (2 L+R)}{2 R}\;,
\notag\\
&=\frac{4\pi k (c_\mathrm{out}^\mathrm{(0)}-c_0)^2 L^2}{\Lambda_\mathrm{d}2}\left(L^2 R + LR^2/2\right)\;.
\end{align}
Now, we still need to consider the modified outside concentration 
\begin{align}
c_\mathrm{out} = \frac{e^{-\frac{R}{L}} \left(c_{\text{in}}^{\text{(0)}}-c_0\right) \left(L-R
   \coth \left(\frac{R}{L}\right)\right)}{L} +c_0\;.
\end{align}
We then get
\begin{align}
F_\mathrm{react}^\mathrm{out} &=\frac{4\pi k (c_\mathrm{in}^\mathrm{(0)}-c_0)^2 L^2}{\Lambda_\mathrm{d}2}\left(\frac{e^{-\frac{R}{L}} \left(L-R
   \coth \left(\frac{R}{L}\right)\right)}{L}\right)^2\left(L^2 R + LR^2/2\right)\;,
\notag\\
   &\approx \frac{2\pi k (c_\mathrm{in}^\mathrm{(0)}-c_0)^2}{\Lambda_\mathrm{d}9} R^5\;,
\end{align}
where we expanded for small droplet radii in the last line.
The total energy is thus given by 
\begin{align}
F_\mathrm{react}^\mathrm{tot} = \frac{4}{15}\frac{ \pi k (c_\mathrm{in}^\mathrm{(0)}-c_0)^2}{\Lambda_\mathrm{d}} R^5\;.
\label{eq:energy-total-3d}
\end{align}

\subsubsection{Approximation for sessile droplets}
We now want to use the two expressions above to estimate the reactive energy $F_\mathrm{react}^\text{sph.cap}$ for sessile droplets forming a spherical cap on a wall.
The volume of a spherical cap translates to its radius as
\begin{subequations}
\begin{align}
    R^{2\mathrm{D}}(V) &= \sqrt{2V^{2\mathrm{D}}/\pi}\;,\\
    R^{3\mathrm{D}}(V) &= ({3V^{3\mathrm{D}}/(2\pi)})^\frac{1}{3}\;.
\end{align}
\end{subequations}
We can now insert these identities into the full expressions for the energy of a sessile droplet on an attractive wall to obtain an upper estimate for the energy of a neutral spherical cap.
Note that we need to correct this energy by a factor of $\frac{1}{2}$ since we calculated the energy of a full droplets.
Hence,
\begin{align}
    F_\mathrm{react}^\text{sph.cap} \approx \frac{1}{2}F_\mathrm{react}^\mathrm{tot}(R(V))\;,
\end{align}
where $F_\mathrm{react}^\mathrm{tot}$ is given by \Eqref{eq:energy-total-2d} for 2D and by \Eqref{eq:energy-total-3d} for 3D.

\subsection{Energy contributions for two droplets}
For the two-droplet scenarios, the energy scaling with the volume remains unchanged.
For the interface contribution, we now need to consider two contributions with half of the total volume. 
We thus get
\begin{subequations}
\begin{align}
    E^{2D}_\mathrm{int} &= \sqrt{2}\gamma_\mathrm{ds}\sqrt{2V(2\vartheta-\sin(2\vartheta))} \;,\\ 
    E^{3D}_\mathrm{int} &= \sqrt[3]{2}4\pi \left(\frac{3V^\mathrm{3D}}{\pi(2+\cos(\vartheta))(1-\cos(\vartheta)^2)}\right)^{(2/3)}\sin^4(\vartheta/2)\;. 
\end{align}
\end{subequations}
The reactive energy can be calculated as follows
\begin{subequations}
\begin{align}
    F^\mathrm{2D}_\mathrm{react} &= b_\mathrm{2D} k V_1^2 +  b_\mathrm{2D} k V_2^2 =  2 b_\mathrm{2D} k \left(\frac{V}{2}\right)^2 = \frac{1}{2} b k V^2\;,\\
    F^\mathrm{3D}_\mathrm{react} &= b_\mathrm{3D} k V_1^{(5/3)} +  b_\mathrm{3D} k V_2^{(5/3)} =  2 b_\mathrm{3D} k \left(\frac{V}{2}\right)^{(5/3)} = \frac{\sqrt[3]{2}}{2} b_\mathrm{3D}  k V^{(5/3)} \;.
\end{align}
\end{subequations}

\subsection{Energy of two interacting chemically active droplets}
We next consider an additional energy contribution to describe the repulsion between the two adjacent droplets. 
To keep it simple, we approximate the two droplets as point charges with separation $l_\mathrm{2D}=2R_1 \sin(\vartheta)=\sqrt{2}R \sin(\vartheta)$, and $l_\mathrm{3D}=2R_1 \sin(\vartheta)=(2/\sqrt[3]{2})R \sin(\vartheta)$. 
The total charge inside each droplet is given by $q=\int(c_0-c(\mathbf{r}))dV\approx (c_0-\cIn)V/2$ and the potential is
\begin{subequations}
\begin{align}
    \Psi_\mathrm{2D} &= \frac{1}{2\pi}q \log\left(|x|^{-1}\right)\;,\\
    \Psi_\mathrm{3D} &= \frac{1}{4\pi}q (|x|)^{-1}\;,
\end{align}
\end{subequations}
where we assume that the connection between the two droplet centers is along the $x$-axis and one charge is located at the origin.
The associated energy to bring a charge from the boundary of the system to the position $l$ is then in 2D given by 
\begin{align}
    F^\mathrm{2D}_\mathrm{rep} &= \frac{k}{2\mobD}\int_{L_\mathrm{sys}}^{l} q\delta(-x)\Psi \text{d}x\;,
\notag\\
                   &=\frac{k(\cIn-c_0)^2 V^2}{16 \pi \mobD} \log(L_\mathrm{sys}/l)\;,
\notag\\
                   &=\frac{k(\cIn-c_0)^2}{16 \pi \mobD} \log\left(L_\mathrm{sys}/\left(\sqrt{2}\sqrt{2V^\mathrm{2D}/(2\vartheta - \sin(2\vartheta)} \sin(\vartheta)\right)\right)V^2   \;.
\end{align}
In three dimensions, we can safely assume an infinite system, such that the reference point of the potential vanishes.
We consequently get
\begin{align}
        F^\mathrm{3D}_\mathrm{rep} &=\frac{k(\cIn-c_0)^2}{32 \pi \mobD} \left((2/\sqrt[3]{2})\left(\frac{3V^\mathrm{3D}}{\pi(2+\cos(\vartheta)(1-\cos(\vartheta)^2)}\right)^{(2/3)} \sin(\vartheta)\right)^{-1}V^2   \;.
\end{align}

\bibliography{lit_jcp}